\input harvmac
\input epsf
\noblackbox
%
\def\ul#1{$\underline{#1}$}
\def\bE#1{{E^{(#1)}}}
\def\IZ{{\bf Z}}\def\IC{{\bf C}}
\def\IR{\relax{\rm I\kern-.18em R}}
\def\la{\lambda}
\def\fc#1#2{{#1\over#2}}

\def\subsubsec#1{\noindent {\it #1} \br}

\def\gn{\Gamma_0}
\def\bgn{{\bar \Gamma}_0}

\def\abstract#1{
\vskip .5in\vfil\centerline
{\bf Abstract}\penalty1000
{{\smallskip\ifx\answ\bigans\leftskip 2pc \rightskip 2pc
\else\leftskip 5pc \rightskip 5pc\fi
\noindent\abstractfont \baselineskip=12pt
{#1} \smallskip}}
\penalty-1000}
\def\us#1{\underline{#1}}
\def\hth/#1#2#3#4#5#6#7{{\tt hep-th/#1#2#3#4#5#6#7}}
\def\nup#1({Nucl.\ Phys.\ $\us {B#1}$\ (}
\def\plt#1({Phys.\ Lett.\ $\us  {B#1}$\ (}
\def\cmp#1({Comm.\ Math.\ Phys.\ $\us  {#1}$\ (}
\def\prp#1({Phys.\ Rep.\ $\us  {#1}$\ (}
\def\prl#1({Phys.\ Rev.\ Lett.\ $\us  {#1}$\ (}
\def\prv#1({Phys.\ Rev.\ $\us  {#1}$\ (}
\def\mpl#1({Mod.\ Phys.\ Let.\ $\us  {A#1}$\ (}
\def\ijmp#1({Int.\ J.\ Mod.\ Phys.\ $\us{A#1}$\ (}
\def\br{\hfill\break}\def\ni{\noindent}

\def\cx#1{{\cal #1}}\def\IP{{\bf P}}

\vskip-2cm
\Title{\vbox{
\rightline{\vbox{\baselineskip12pt\hbox{CERN-TH/98-166}
                                  \hbox{EFI-98-20}
                                  \hbox{NSF-ITP-98-67}                                  
                                  \hbox{hep-th/9805189}}}}}
{On Type IIB Vacua With Varying}
\vskip-1cm\centerline{{\titlefont Coupling Constant}}
\vskip 0.3cm
\centerline{P. Berglund$^a$, A. Klemm$^{a,b}$, P. Mayr$^{a,c}$ and S. 
Theisen$^{a,d}$}
\vskip 0.6cm
\centerline{$^a$ \it Institute for Theoretical Physics, 
University of California, Santa Barbara, CA 93106, USA}
\vskip 0.0cm
\centerline{$^b$ \it 
Enrico Fermi Institute, University of Chicago, Chicago, IL 60637, USA} 
\vskip 0.0cm\centerline{$^c$ \it Theory Division, CERN, 1211 Geneva 23, 
Switzerland}
\vskip 0.0cm\centerline{$^d$ \it Sektion Physik, Universit\"at M\"unchen, 
D-80333 
Munich, Germany}
\vskip 0.3cm
\abstract{We describe type IIB compactifications with varying
coupling constant in $d=6,7,8,9$ dimensions, where part of the 
ten-dimensional $SL(2,\IZ)$ symmetry is broken by a background
with $\Gamma_0(n)$ or $\Gamma(n)$ monodromy for $n=2,3,4$.
This extends the known class of F-theory vacua to theories which
are dual to heterotic compactifications with reduced rank. On 
compactifying on a further torus, we obtain a description of 
the heterotic moduli space of $G$ bundles over elliptically fibered 
manifolds without vector structure in terms of complex geometries.
}

\Date{\vbox{\hbox{\sl {May 1998}}
}}
\goodbreak

\parskip=4pt plus 15pt minus 1pt
\baselineskip=15pt plus 2pt minus 1pt

\newsec{Introduction}
F-theory \ref\vafaf{C. Vafa, \nup 469 (1996) 403.} 
provides an interesting dual description 
of heterotic strings in various dimensions. Amongst the most
promising aspects is the fact that some non-geometric data of the
heterotic string are included in the geometric description 
of the F-theory compactification. 
A large class of F-theory/heterotic duals has been described in
\ref\mvi{D. R. Morrison and C. Vafa, \nup 473 (1996) 74.}%
\ref\mvii{D. R. Morrison and C. Vafa, \nup 476 (1996) 437.}%
\ref\cf{P. Candelas and A. Font, \nup 511 (1998) 296.}%
\ref\bv{M. Bershadsky et al., \nup 481 (1996) 215.}%
\ref\esp{P. Aspinwall and D. R. Morrison, \nup 503 (1997) 533.}\foot{For a 
detailed one-loop comparison in 8d, see \ref\ls{W.\ Lerche, S.\ 
Stieberger, {\sl Prepotential, Mirror Map and F-Theory on K3},
hep-th/9804176.}.}. 
The purpose of this note is to extend this dictionary to a large class of 
heterotic theories with extra $U(1)$ backgrounds. This includes
CHL type of vacua \ref\chl{S. Chaudhuri, G. Hockney and J. D. Lykken,
\prl 75 (1995) 2264.} in various dimensions as well as $U(1)$ 
instanton backgrounds. 

Let us recall the crucial ingredients of the construction in 
\vafaf. Type IIB theory in ten dimensions has a conjectured 
strong/weak coupling symmetry $\gn=SL(2,\IZ)$. $\bgn=PSL(2,\IZ)$  
acts on the complex dilaton field
$\tau=\phi+i\exp(\tilde{\phi})$ \ref\hullt{C.M. Hull and P.K. Townsend,
\nup 438 (1995) 109.} by fractional linear transformations. 
$\gn$ acts also non-trivially on the 
two-form potentials $B^{NS}$ and $B^{RR}$ of the theory while
leaving the other bosonic fields invariant. The idea of \vafaf\
is to interpret the complex parameter $\tau$ as the modulus of a complex
torus $E$ and consider type IIB compactifications on a base 
manifold $B$ with varying $\tau$. Since the monodromy group of 
the moduli space of the torus $E$ is generically $\gn$, 
$\tau$ will jump by an $\bgn$ transformation 
at those points $b\subset B$ on the base, where the torus $E$ degenerates. 
Due to the $\gn$ symmetry of type IIB this is nevertheless a
valid vacuum. In physical terms, the points where $\tau\to i\infty$ 
are to be identified as $D7$ branes of the type IIB theory.
The type IIB equations of motion require the total manifold $X_n$
of the fiber $E$ together with the $n-1$ dimensional base $B_{n-1}$ 
to be a Calabi--Yau manifold. By construction, it has an elliptic
fibration $\pi:X_n\to B_{n-1}$.

Note that generically, the ten-dimensional $\gn$ symmetry
is completely broken by the non-trivial vacuum due to the fact
that the monodromy of $E$ generates all of $\gn$. In particular,
the BPS states, in a given $\gn$ orbit of BPS states, of 
the ten-dimensional type IIB theory are transformed into each other
by the action of the $\gn$ monodromy.

It is natural to extend this construction by compactifications
which do not generate all of the $\bgn$ transformations acting
on $\tau$. This does not necessarily mean that there is a left-over
symmetry group of the lower-dimensional theory corresponding to
a non-trivial commutant $H$ of a subgroup $\Gamma\subseteq\gn$.
Instead, we consider the case where $\Gamma$ is a finite
index subgroup of $\gn$ with trivial commutant $H$. Upon 
compactification, each $\gn$ orbit of BPS states of the 
ten-dimensional type IIB theory splits  into several
orbits of the coset $\gn/\Gamma$ 
leading to a very different spectrum in the
lower dimensional theory. This is comparable to the situation in
$N=4$ and $N=2$ Super-Yang-Mills theories in four dimensions: 
in the $N=4$ theory with a $\gn$ strong-weak coupling duality, 
magnetic and electric degrees
of freedom can be equivalent due to the fact that they are living
in a single $\gn$ multiplet. In the $N=2$ theory with 
a $\Gamma \subset \gn$ symmetry, there are 
two very different kinds of short multiplets corresponding to 
the vector multiplets and the monopole hypermultiplets. The
charges of the two types of multiplets are related by the ``missing''
transformations\foot{E.g. for pure N=2 $SU(2)$ theory the subgroup 
is $\Gamma_0(2)$.} in $\gn/\Gamma$. Note that the distinct
BPS spectra of the theories with different symmetry groups 
rule out a continuous interpolation from one theory to the
other at finite distance in moduli space.

What we find is that the type IIB compactifications
with restricted monodromy on $\tau$ lead to moduli spaces
describing Calabi--Yau manifolds with frozen moduli. These
restricted geometric moduli spaces, which have a natural 
interpretation in terms of toric geometry as we will see, 
correspond to heterotic compactifications with frozen moduli 
via F-theory/heterotic duality. In the heterotic theory 
the freezing can be either due to a CHL type modding or due to
$U(1)$ instanton backgrounds. A type IIB interpretation 
for the freezing in the case of monodromy group $\Gamma_0(2)$
has been described in six- and eight-dimensional
orientifold constructions:
it corresponds to the choice of a non-trivial $\IZ_2$ background for
the Neveu-Schwarz two-form $B_{\mu\nu}$. In particular,
non-triviality of the $B_{\mu\nu}$ background is related to 
the absence of vector structure in the type I $SO(32)$ string 
\ref\sense{A. Sen and S. Sethi, \nup 499 (1997) 45.} 
which in turn corresponds to 
a reduced rank compactification described by a CHL vacuum
\ref\witwov{E. Witten, {\it
Toroidal compactification without vector structure}, hep-th/9712028.}.
The choice of a $\IZ_2$ valued $B$ field is compatible with the 
restricted monodromy group $\Gamma_0(2)$. 
There appears to be one additional non-trivial choice apart from a
$\IZ_2$ valued $B$ field in eight dimensions, namely the 
case $\IZ_4$. The other case that
we discuss, $\Gamma_0(3)$, cannot be pushed to a
higher than seven dimensional vacuum.

The paper is organized as follows. In section~2 we study F-theory
compactifications with the elliptic fiber  of a less generic type,
i.e. with $\Gamma\subset \gn$ monodromy. We then turn to the gauge
enhancements and the relevant dual CHL heterotic theories in section
~3, and in particular when there exists an F-theory limit to eight
dimensions. The heterotic interpretation naturally gives
the moduli space of $G$ bundles without vector structure on elliptic
curves, as discussed in section~4.

While this work was being finished for
publication we received a preprint 
by Bershadsky, Pantev and Sadov 
\ref\bps{M. Bershadsky, T. Pantev and V. Sadov, hep-th/9805056.}, 
which, to a certain extent, overlaps with our results.

\newsec{Type IIB geometries with $\Gamma\subset \gn=SL(2,\IZ)$ 
monodromy}

To define a type IIB compactification with varying coupling constant along
the lines of ref. \vafaf, we have to give an elliptically fibered
Calabi--Yau manifold $X_n$ of complex dimension $n$ together with a section 
$\sigma$,
on which the type IIB string is compactified. Let $E$ denote the elliptic
fiber and $B_{n-1}$ the base of the fibration $X\to B$. A model
for $E$ and $X_n$ can be given in terms of (an intersection of) 
hypersurfaces in a projective space (or generalizations thereof).  
As we will explain in a moment, the model for $E$ will determine 
the monodromy group $\bar \Gamma \subset \bar\gn$ acting on 
the complex structure $\tau$ of $E$
in the fibration. In particular, $\Gamma$ being contained in
$\Gamma_0(n)$  
is related to the presence of shift symmetries of finite order 
of the elliptic fiber $E$. Specifically\foot{Some facts
about $\Gamma_0(n)$ and $\Gamma(n)$ are collected in Appendix A.},
$\Gamma_0(n)$ ($\Gamma(n)$)
preserves $n$ $(n^2)$ points of order $n$ on $E$.

\subsec{Fiber geometry and  fibrations over $\IP^1$}
To be concrete, we will consider models of elliptic curves
$E$ given as the zero of one or two polynomials of appropriate degree in
weighted projective space and K3 fibrations with $E$ as the fiber.
The K3 case contains all the 
essential features also for the case of higher-dimensional
Calabi--Yau manifolds with restricted monodromies, which can
be treated very similarly.

\ni
\subsubsec{Generic case: $\gn$ monodromy}
Let us consider a model  of a generic elliptic curve
$E$ given as the zero of a polynomial of degree six in weighted 
projective space $\IP^2_{1,2,3}$.
\def\bS#1{{\cal S}_{#1}}        
\eqn\toria{
\bE1:\qquad
y^2+x^3+z^6+\mu x z^4=0,
\qquad (z,x,y) \in \IP^2_{1,2,3}\ .}
To get a K3 fibration with $\bE1$ as the generic fiber, we 
fiber $\bE1$ over a base $\IP^1$ by making the coefficients
of the monomials in $\bE1$ functions\foot{More precisely sections of
$\cx O_{\IP^1}(l)$.} of the base coordinates $(s,t)$ of $\IP^1$ in a way 
compatible with the Calabi--Yau condition. In this way we get 
a K3 \def\bX#1#2{{ X_{#1}^{(#2)}}}
\eqn\kthreea{
\bX{2}{1}:\qquad y^2+x^3+xz^4f_8+z^6f_{12}=0\ ,
}
where $f_l$ are degree $l$ polynomials in the base variables
$(s,t)$. The elliptic fibration of $\bX{2}{1}$ has the generic
$\gn$ monodromy, and F-theory compactified on $\bX{2}{1}$ is
dual to the standard heterotic compactification on $T^2$ in 
eight dimensions with
gauge group $G$ of rank 18 \vafaf. The zeros of the discriminant 
$\delta=4f_8^3+27f_{12}^2=0$ are identified with the locations of D7 branes
on the $\IP^1$ base. For generic $f_8$ and $f_{12}$ there are
24 simple zeros corresponding to singular elliptic fibers. The
total manifold is nevertheless smooth since a singularity in the
K3 would correspond to a zero of order $\geq 2$ in $\delta$.
The independent number of D7 brane positions 
determining the rank of $G$ in the F-theory compactification
is given by the $13+9-4=18$ independent parameters in \kthreea,
where we had to subtract 3 parameters for the reparametrizations
of the base coordinates and one overall scaling.\br\ni

\subsubsec{Type IIB backgrounds with $\Gamma_0(2)$ and $\Gamma(2)$ monodromy}

Instead of the model \toria\ for $E$, we consider a degree
four polynomial in $\IP^2_{1,1,2}$. The generic polynomial 
has $9$ terms, but the $8$ parameters in the redefinitions of $(x,y,z)$ 
compatible
with the weights allow us to 
choose a simple form for the defining equation  
\eqn\torib{\bE2:\qquad 
y^2+x^4 + x^2 z^2 \mu + z^4=0\ , \qquad (x,z,y) \in \IP^2_{1,1,2}\ .}

In order to remove the dependence on the choice of representative, 
the monodromy subgroup is usually obtained by considering
\eqn\jgammatwo{\hat\jmath(\mu)={(12+\mu^2)^3\over 108 (\mu^2-4)^2}= j(\tau),}  
where $\hat\jmath(\mu)$ is invariant under changes on the coordinates $(x,y,z)$ 
and $j(\tau)$ is the $\bgn$ invariant function.
Eq. \jgammatwo~does not quite imply that $\mu(\tau)$ is 
$\bgn$ invariant. 
The  involutions 
$m:\mu\mapsto -\mu$ and $n:\mu\mapsto {{2 \mu + 12}\over{\mu-2}}$
leave $\hat\jmath(\mu)$ invariant. They satisfy $(mn)^3=1$ and thus
generate $S_3$, which can be identified with the group permuting the roots of 
\jgammatwo, read as a sixth order polynomial in $\mu$. 
The monodromy group is then $\Gamma(2)$ with 
$\bgn/{\bar\Gamma}(2)\simeq S_3$. 
   
 
However, given the preferred representation \torib\ there is an easier
way to see the monodromy group by noting that $\bE2$ enjoys two 
$\IZ_2$ symmetries generated by\foot{Note that there could be 
additional polynomials in the 
defining equation, which we have removed by variable redefinitions
of the ambient space, that do  not respect these symmetries as e.g. 
a $x z^3$ term.} 
\eqn\shiftone{
\bS1:(y,x,z)\to (-y,z,x), \qquad \bS2:(y,x,z)\to (-y,-x,z).}
It can be verified that the symmetries $\bS i$ act without fixed 
points and leave the holomorphic one-form invariant. 
In general, a
symmetry of the elliptic curve always translates to a symmetry 
of the lattice $\Lambda$ spanned by the periods, where $E$ is defined 
as $E={\bf C} /\Lambda$. In particular, fixpoint free $\IZ_n$ symmetries 
must be order $n$ shift symmetries on $E$. The presence of an order $n$ shift 
implies that there are $n$ points in a finer 
lattice\foot{From which an isogenous curve $E'$ can be defined},
so-called points of order $n$, which are equivalent to 
the origin up to a $\IZ_n$ shift. Clearly, the monodromy does not change the 
symmetries and preserves 
the non-trivial order $n$ points. Therefore the  monodromy group must 
be a subgroup of $\Gamma_0(n)$, as further explained in Appendix A.  
Similarly, two independent $\IZ_n$ shifts, corresponding to a finer lattice
in the two directions of $\Lambda$,  imply the existence of 
$n^2$ such points and the monodromy group 
$\Gamma(n)$. Below we will fiber tori with known modular groups
and by fibering we can only lower the symmetries to a subgroup.
Thus, the disadvantage of the second method that one
might not find all symmetries, is not severe.

To get fibrations with restricted monodromy, we take the 
shift invariant form $\bE2$ and make the coefficients again vary
with the coordinates $(s,t)$ on the base $\IP^1$ in a way compatible
with the Calabi--Yau condition:
\eqn\kthreeb{
\bX{2}{2}:\qquad y^2+x^4+x^2z^2f_4+z^4f_8=0\ .}
Since one $\IZ_2$ 
shift is broken, the monodromy of the fibration is $\Gamma_0(2)$
rather than $\Gamma(2)$.

As before, we can determine the rank of the gauge group $G$ by
counting the number of independent parameters in 
the polynomial; this time
we get $5+9-4=10$. Thus, compactification of F-theory 
on $\bX{2}{2}$ is dual to heterotic compactification on $T^2$
with a rank reduction of $-8$ as compared to the standard heterotic
compactification.
This is characteristic of the original CHL compactification,
which can be understood as a compactification on $T^2$ which 
combines a half-shift on a circle of $T^2$  with an exchange of 
the two $E_8$ factors in the $E_8\times E_8$ heterotic string or
the exchange of two $SO(16)$ factors in the $SO(16)\times SO(16)\subset
SO(32)$ of the $SO(32)$ heterotic string 
\ref\cp{S. Chaudhuri and J. Polchinski, \prv (1995) 7168.}
\foot{A more detailed analysis of this case has appeared in 
\ref\lsmt{W. Lerche, R. Minasian, C. Schweigert and S. Theisen,
\plt 424 (1998).}.}.
As for the D7 brane
locations, the discriminant of the elliptic fibration $\bX{2}{2}$
is given by $\delta=f_8(4f_8-f_4^2)^2$. 
Thus, we have 8 simple zeros which correspond to eight 
dynamical D7 branes as well as 8 double zeros corresponding to 
$A_1$ singularities of the manifold $\bX{2}{2}$.

The $j$-function of the elliptic fibration $\bX{2}{2}$ is 
\eqn\jagain{
j = \fc{1}{108} \fc{(f_4^2+12 f_8)^3}{f_8 (-f_4^2+4f_8)^2} \ .}
As explained in Appendix A, the $\Gamma_0(2)$ symmetry
implies the presence of a second section. This can be verified 
from the factorization of the Weierstrass form 
$y^2=4 x^3-g_2 x-g_3$
\eqn\wsb{
y^2=( 4 \, x + {4 f_4\over \sqrt{3} })\, (x^2-{f_4\over\sqrt{3} }\, x-
3\, f_8 +{f_4^2\over 12})}
We will justify the
identification of F-theory on $\bX{2}{2}$ and the existence of
a eight-dimensional limit in the next section, where we explore the
geometrical properties of $\bX{2}{2}$.

We can restrict the monodromy further to $\Gamma(2)\subset \Gamma_0(2)$
by choosing $f_8=\hat f_4^2$. We then obtain a rank $10-4=6$ 
theory which should correspond to a heterotic theory with rank
reduced by $-12$. The corresponding heterotic CHL vacua in six dimensions
have also been constructed in \ref\cl{S. Chaudhuri and D. Lowe, 
\nup 459 (1996) 113.}. The singularity structure is $(A_1)^{12}$. Moreover, 
the Weierstrass form
\wsb\ factorizes further corresponding to four global sections of
the elliptic fibration.

\subsubsec{Type IIB backgrounds with $\Gamma_0(3)$ and $\Gamma(3)$ monodromy}
In order to get a monodromy restricted to $\Gamma_0(3)$, we consider
an elliptic curve $\bE3$ given as a degree three
polynomial in $\IP^2_{1,1,1}$:
\eqn\toric{
\bE3:\qquad y^3+x^3+z^3+\mu yxz=0,\qquad
(y,x,z) \in \IP^2_{1,1,1} .}
This torus enjoys two $\IZ_3$ shift symmetries generated by 
\eqn\shifttwo{
\bS1:(y,x,z)\to (x,z,y),\qquad \bS2:(y,x,z)\to (y,\la x,\la^2 z)\,,
\quad\la=e^{2\pi i\over 3}\,.}
The monodromy group in this case is $\Gamma(3)$ with 
$\bgn/{\bar\Gamma}(3)\simeq A_4$.

Similarly as before, we get a K3 with $\bE3$ as the generic
elliptic fiber:
\eqn\kthreec{
\bX{2}{3}:\qquad y^3+x^3+yxzf_2+z^3f_6=0\ .}
For reasons similar to above, the 
monodromy of the fibration is $\Gamma_0(3)$.
Counting the number of independent parameters 
we get $3+7-4=6$ for the rank of the gauge group. 
We will identify this theory in the next section with a CHL
type construction, where we first compactify the heterotic $SO(32)$ string
on $T^2$ to get an $SO(36)$ gauge group, and then combine a $\IZ_3$ shift
on a further $T^2$ with a permutation of the three $SO(12)$ factors
in $SO(12)^3\subset SO(36)$ to get a six-dimensional vacuum 
with reduction of rank $-12$ as compared to the standard compactification \cl. 
The discriminant has the form
$\delta=f_6(27f_6+f_2^3)^3$. Thus, in this case 
we have 6 simple zeros corresponding to six
dynamical D7 branes as well as 6 triple zeros corresponding to 
$A_2$ singularities of $\bX{2}{3}$.

If we further restrict $f_6$ to be the cube of a degree two 
polynomial $\hat f_2$, we obtain a elliptic fibration with $\Gamma(3)$
monodromy. The singularity structure is $(A_2)^8$ and the number of remaining
parameters is $3+3-4=2$, corresponding to a rank $-16$ reduction
on the heterotic side. A CHL vacuum with this properties has been given in
\ref\clii{S. Chaudhuri and D. Lowe, \nup 469 (1996) 21.}

We can also look for points in the moduli space where the type IIB coupling
$\tau$ becomes constant as in ref. \ref\senf{A. Sen, \nup 475 (1996) 562.}. 
In particular such a configuration would be a possible candidate for an
orientifold vacuum of type IIB. The $j$-function of the elliptic fibration 
of $\bX{2}{3}$ reads
\eqn\joncemore{
j=-\fc {1}{1728} \fc{f_2^3(-216f_6+f_2^3)^3}{f_6(27f_6+f_2^3)^3}}
A constant coupling limit which intersects weak type IIB coupling requires
$f_6\sim f_2^3$. However, for this choice $\bX{2}{3}$ acquires
a $\hat{E}_8\times\hat{E}_8$ singularity which has no valid interpretation
as a weakly coupled orientifold.

\subsubsec{Type IIB backgrounds with $\Gamma_0(4)$ and $\Gamma(4)$ monodromy}
Restricted $\Gamma_0(4)$ monodromy can be obtained by choosing a
complete intersection in $\IP^3$ as a model for the elliptic fiber $E$:
\eqn\torid{
\bE4:\qquad u^2+v^2+\mu xz=0\ ,\ \ x^2+z^2+\mu uv=0\ ,\qquad
(u,v,x,z) \in \IP^3 .}
This torus has two $\IZ_4$ shift symmetries generated by 
\eqn\shiftthree{
\bS1:(u,v,x,z)\to (x,z,v,u),\qquad
\bS2:(u,v,x,z)\to (\la u, \la^{-1} v,\la^2 x,z), \ \
\la =e^{\fc{2\pi i}{4}}.}
The monodromy group is $\Gamma(4)$ with $\bgn/{\bar\Gamma}(4)\simeq S_4$.
As before, we can find a K3 with restricted monodromy $\Gamma_0(4)$ with
$\bE4$ as the generic elliptic fiber:
\eqn\kthreed{
\bX{2}{4}:\qquad u^2+v^2+xzf_2=0\ ,\ \ x^2+z^2f_4+uv=0 \ .}
Counting the number of independent parameters  
we get $3+5-4=4$ for the rank of the gauge group. 
We suggest to identify this theory with a CHL
type construction in eight dimensions, 
where the four factors $SO(8)^4\subset SO(32)$ are permuted together
with a $\IZ_4$ shift on $T^2$ \cl. 
The discriminant of the elliptic fibration $\bX{2}{4}$
is $\delta=f_2^2f_4(f_4-\fc{1}{16}f_2^2)^4$. Thus, in this case 
we have 4 simple zeros corresponding to four
dynamical D7 branes as well as 2 double zeros and 4 zeroes of order
4 corresponding to $A_1^2\times A_3^4$ 
singularities of $\bX{2}{4}$.

The $j$-function for the elliptic fibration $\bX{2}{4}$ is
\eqn\jlast{
j = \fc{1}{1728}\fc{(f_2^4+224 f_4 f_2^2+256 f_4^2)^3}
{f_2^2 f_4 (f_2^2-16 f_4)^4} \ .}
It can be verified that the Weierstrass form factorizes indicating
a second section as expected. 

Restricting $f_4=\hat f_2^2$ we get an elliptic fibration with 
$(A_1)^4\times (A_3)^4$ singularity and $6-4=2$ parameters.
Moreover, for $\hat f_2 \sim f_2$ we get a configuration with
constant coupling. As in the case of $\Gamma(3)$ however, 
the singularity structure in this limit is $\hat{E}_8 \times
\hat{E}_8$ which cannot be interpreted as an orientifold. $\Gamma(4)$ can
only be recovered if $f_4=f_1^4$ and $f_2=\tilde f_1^2$. In this case
we have $2+2-4=0$ no moduli left.

The coordinates for the four elliptic fibrations $\bX{2}{1}$, $\bX{2}{2}$,
$\bX{2}{3}$, $\bX{2}{4}$ are defined by the equivalence classes 
under the $(\IC^*)^2$ actions
\eqn\scalings{\eqalign{
(x,y,z,s,t)&\sim (\lambda^2 x, \lambda^3 y , \lambda \rho^{-2} z, \rho s, \rho 
t) \cr
(x,y,z,s,t)&\sim (\lambda x, \lambda^2 y , \lambda \rho^{-2} z, \rho s, \rho 
t)\cr
(x,y,z,s,t)&\sim (\lambda x, \lambda y , \lambda \rho^{-2} z, \rho s, \rho t)\cr
(u,v,x,z,s,t)&\sim (\lambda u, \lambda v , \lambda x,\lambda \rho^{-2} z, \rho 
s, \rho t)\ .}}
To define smooth coordinates the loci $s=t=0$ and $y=x=z=0$ for the 
first three cases and $u=v=x=z=0$ for the last case
have to be excluded, since the dimension of the orbits of the 
$\IC^*$ actions changes there.  

For an 8d dimensional F-theory interpretation we must be able to declare 
a $\IP^1$ section of the elliptic fibration as the space the theory 
actually lives on. For a generic choice of the monomials in the 
fibrations~\toria,\torib,\toric\ and \torid~the unique section is the 
intersection of the 
$z=0$ divisor with the constraint(s) and one finds, in view of the 
above equivalence relations, that the divisor is 
described by $(s,t)$, the homogenous $\IP^1$ coordinates, 
times\foot{$'$ means correlated signs.}  
\eqn\multisec{\eqalign{
y^2&=x^3,\qquad \qquad \qquad \! 1\  {\rm point}  \ [x,x] \cr
y^2&=x^4,\qquad \qquad \qquad \! 2\   {\rm points} \ [x,\pm x] \cr
y^3&=x^3,\qquad \qquad \qquad \! 3\  {\rm points} \ [x,e^{2 \pi i n\over 3} x],\ 
\ n=0,1,2\cr 
u^2&=v^2,\quad x^2=uv,\quad 4\ {\rm points} \ [\pm' u,\pm' v,\pm x].}}
For arbitrary choice of the monomials in the constraints 
the points in the latter three cases are permuted by monodromy in the $(s,t)$ 
coordinates and the sections are referred to as multi-sections. 
The K\"ahler classes of these multi-sections descend from only one 
K\"ahler class of the ambient space, the one associated to the $z=0$ 
divisor, and are therefore not independent but 
locked together\foot{In general this phenomenon manifests itself in 
the toric polyheder diagrams via the occurrance of points of the corresponding 
multiplicity; c.f. section 3.4.}. In the small fiber 
$F$-theory limit the sections are identified and the distinction 
between the four above cases disappears.

New 8d theories exist only if one can restrict the complex 
deformations in such a way that one gets, e.g. by splitting of the 
multi-sections, independent global sections, one of which, $\sigma_0$, 
will become the compactification space. In the $F$-theory limit the 
remaining parts of the multi-sections can be shrunk together with the fiber 
and other K\"ahler classes not intersecting $\sigma_0$. 
This leads to a rank reduction as will be explained 
further below.

\subsec{Rank reduction in F-theory}
In the F-theory picture the rank of the gauge group in eight
dimensions is usually 18. This is because one needs
24 D7 branes to make the transverse space compact, however
only 18 D7 branes 
can be chosen to have the same $(p,q)$
charge under the $\gn$ of type IIB and are therefore independent \vafaf.
The D7 branes are located at points on the section
of the elliptic fibration which is isomorphic to the $\IP^1$
base of the K3 fibration. This is the space the type IIB string is 
compactified on.
To reduce the rank, we somehow need to get
rid of some of the D7 branes. 
It turns out that the F-theory geometries
with restricted monodromy find a clever way to achieve this:
we will find that there are two sections, where on the first one
we have the usual fiber singularities corresponding to a reduced
rank 
gauge group of the type IIB theory in eight dimensions, whereas
on the other section we have singularities which account for the
missing rank in the eight-dimensional theory. The first section 
is the one on which type IIB theory is compactified to eight dimensions
while we have to contract the second section. 
So the rank reduction is due to the fact that the 
type IIB theory sees only the D7 branes on the section on which it
is compactified; these generate the reduced rank gauge group present 
in eight dimensions.

To demonstrate this mechanism we can follow a deformation of the
generic K3 surface to the one relevant for the $\Gamma_0(2)$ 
theory dual to the heterotic CHL vacuum.
We start with the generic non-singular K3 surface
$\IP^{3}_{1,1,4,6}[12]$ with 
generic elliptic fiber $\IP^{2}_{1,2,3}[6]$:
\eqn\ell{
p_A=y^2+x^3+xz^4f_8+z^6g_{12}\ ,
}
where $f_8$ and $g_{12}$ are polynomials of degree $8$ and $12$
in the base variables $s,t$, respectively. We have
18 independent complex deformations in $p_A$ and two
scaling actions $(y,x,z,s,t)\to(\la^3 y,\la^2 x,\la\mu^{-2} z,\mu s ,\mu t)$
representing the two K\"ahler classes generated by the fiber
and a section. 

In fact, \ell\ is a
special choice of representative for the generic polynomial,
due to the freedom to redefine variables compatible
with the weights. A different representative is
\eqn\ellp{
y^2+x^3+xz^4f_8+yz^3g_6+x^2z^2f_4\ .
}
By changing variables, \ellp\ can be put into the form \ell\
up to the fact that we get only 12 parameters for the 
polynomial $g_{12}$, that is \ellp\ describes a special codimension one
subspace of the deformation \ell. The special property of
\ellp\ is that there is a bi-section for the
elliptic fibration, $\tilde{\sigma}:x=y=0$, in addition to the canonical
section $\sigma:z=0$. We can blow up this point by introducing
new variables
\eqn\bup{
x=xu\ ,\qquad y=yu\ ,
}
and get, 

\eqn\pebe{\eqalign{
p_B&=\ uy^2+u^2x^3+xz^4f_8+yz^3g_6+ux^2z^2f_4
\ .}}
$X_B$ is a K3 surface of new algebraic type with Picard lattice
of rank 3.
The new K\"ahler class is related to the new holomorphic elliptic 
curve of type $\IP^2_{1,1,2}[4]$ with variables $(y,z,u)\sim(\la y,\la 
z,\la^2 u)$ 
in $X_B$. 
Note that the two elliptic fibrations share the section $z=0$.

Up to now, all we have done is to restrict to a certain subspace
in complex structure moduli space
of the elliptic fibration. F-theory on $X_B$ 
does {\it not} give rise to a new eight-dimensional theory since
in the limit of small fiber the moduli spaces of $X_A$ and $X_B$
coincide\foot{In particular, blowing down the original elliptic fiber $\bE1$
in $X_B$
always blows down also the second fiber $\bE2$. Therefore we simply 
end up at a special point in the complex structure moduli space 
of $X_A$ in the zero size limit.}. 
Therefore the deformation from $X_A$ to $X_B$ is a 
new branch only in the seven-dimensional M-theory compactification.
This is related to the fact that the two elliptic fibrations in 
$X_B$ do not yet have restricted monodromy. In particular, the
presence of the term $yz^3g_6$ term  is not compatible with the
shift symmetries of $\torib$. 
Note that 
the original elliptic fibration with coordinates $(y,x,z)$ 
has now {\it two} sections\foot{
In the generic form of $p_B$, the point $y=x=0$ corresponds to
a single point in one sheet of the double cover of the Riemann
sphere and moving
on the base $\IP^1$ it is exchanged with the corresponding point
on the other sheet. For $f_6=0$, $y=x=0$ is a branch point of the
double covering, invariant under the monodromies in the base, 
and $\tilde{\sigma}$ becomes an honest global section.}, 
$\sigma$ and $\tilde{\sigma}$. 
Moreover, $X_B|_{g_6=0}$ has an $(A_1)^8$ singularity on the
section $\tilde{\sigma}$, namely $y=x=f_8=0$. 
\def\xh{\hat{x}}

To provide a single section on which type IIB theory is compactified,
we contract $\tilde{\sigma}$ by blowing up the locus $y=x=0$ by introducing 
the coordinate $\xh$ via
\eqn\blowii{y=y\xh,\ x=x\xh^2\ .} 
After this blow-up, the divisor $x=0$ no 
longer intersects the hypersurface $p_C=0$ corresponding to a blow down 
of $\tilde{\sigma}$. Therefore we can set $x=1$ to get a
K3 surface
\eqn\pece{
p_C=uy^2+u^2\xh^4+yzu\xh st+u\xh^2z^2f_4+z^4f_8\,.}

This K3 surface has two elliptic fibrations with the quartic $\bE2$ 
in $\IP^2_{1,1,2}$ as a fiber model; they have coordinates 
$(\xh,z,y)$ and $(y,z,u)$, respectively. The first fibration has
the restricted $\Gamma_0(2)$ monodromy and describes the F-theory
dual to the CHL vacuum in eight dimensions. The second fibration
is not of the $\Gamma_0(2)$ type\foot{Because of the term $uy^2$.}.
This fibration corresponds to a CHL vacuum in six dimensions with
$F_4\times F_4$ gauge symmetry as we will explain below.

It is easy to keep track of the blow-ups in a toric representation 
of the geometry. To this end read the 
shifted exponents $\nu^{\rm mon}=(n_x,n_y,n_s)$  of every monomial 
$y^{n_y+1}x^{n_x+1}s^{n_s+1}$ in~\ell~as a vector 
in the lattice $\Lambda^*=\IZ^3$ embedded in $N^*=\IR^3$. The Calabi-Yau condition
is equivalent to the requirement that the convex hull of all $\nu^{\rm mon}$
is a reflexive polyhedron 
$\Delta^*={\rm Conv}\{\nu^{\rm mon}\}$. This means that the dual 
polyhedron $\Delta=\{x\in N|\langle x,y\rangle \ge -1,\ 
\forall y\in \Delta^*\}$ has only lattice points in $\Lambda$ as 
vertices. The polyhedron $\Delta={\rm Conv}\{\nu^{\rm coord}\}$ 
defines the ambient space, in our case  $\IP^3_{1,1,4,6}$, 
and the points $\nu^{\rm coord}$ provide the coordinates such that the defining equation, 
in our case~\ell , reads~\ref\batyrev{Victor V. Batyrev, Duke Math. 
Journ {\bf 69} (1993) 349.}\ref\batcox{D.\ Cox, {\sl 
The Homogeneous Coordinate Ring of a Toric Variety}, alg-geom/9210008.} 
\eqn\batcox{p=\sum_{i} a_i \prod_{j} x_j^{\langle \nu_i^{\rm mon},
\nu_j^{\rm coord}\rangle+1}=0\ .}  
Here the sum runs over points $\nu_i^{\rm mon}$ of $\Delta^*$ and the product over
points $\nu^{\rm coord}_j$ of $\Delta$. The fact that for the reflexive pair
$(\Delta,\Delta^*)$ the r\^ole of $\Delta^*$ and $\Delta$ can be 
exchanged in \batcox~is the manifestation of mirror symmetry in Batyrevs 
construction. Generally, every reflexive pair of $d+1$ dimensional polyhedra 
$(\Delta,\Delta^*)$ defines in this way via~\batcox~a 
$d$ dimensional Calabi-Yau manifold $X$
($d=2$ for $K3$) and its mirror\foot{ For K3 rank(Pic) of $X$ is exchanged 
with the dimension of the deformation space of the
mirror~\ref\dolgachev{I.\ V.\  
Dolgachev, {\sl Mirror symmetry for Lattice Polarized K3 Surfaces}, 
alg-geom/9502005.}  only if there is no correction term in eq. (3.1).} $X^*$.  

To be concrete, note that for~\ell~$\Delta={\rm Conv}\{\nu^{\rm coord}\}$ 
with $\nu^{\rm coord}=\{(-1,2,0),$ $(1,-1,0),$ $(-1,-1,0),$ 
$(-1,-1,1),$ $(-1,-1,-1)\}$ providing the coordinates $(x,y,z,s,t)$, 
while the allowed monomials in~\ell~correspond to the points in\hfill\break 
$\Delta^*={\rm Conv}\{(1,0,0),(0,1,0),(-2,-3,6),(-2,-3,-6)\}$. 
Fibration structures can be dectected from the toric 
diagrams by the fact that the reflexive polyhedron of the fiber 
is embedded in the polyhedron of the total space~\ref\klm{A.Klemm, W.Lerche, P.Mayr, 
Phys. Lett. B357 (1995) 313.}\ref\akms{A.C. Avram, 
M. Kreuzer, M. Mandelberg and H. Skarke, \nup 494 (1997) 567.}.
The  elliptic fiber is representated by the 
two-dimensional reflexive sub-polyhedra $\Delta_E$ and 
$\Delta^*_E$ in the $(x,y,0)$-planes of $\Delta$ and $\Delta^*$, 
respectively. $\Delta_E$ is shown in figure 1.A  and all blow-ups, some of 
them were discussed above, can be described by adding points in the $\Delta$-plane. 
The $K3$ polyhedron $\Delta$ is completed by adding $\nu_s=(-1,-1,1)$ 
and $\nu_t=(-1,-1,-1)$ to $\Delta_E$. 

\vskip 0.5cm
{\baselineskip=12pt \sl
\goodbreak\midinsert
\centerline{\epsfxsize 3.5truein\epsfbox{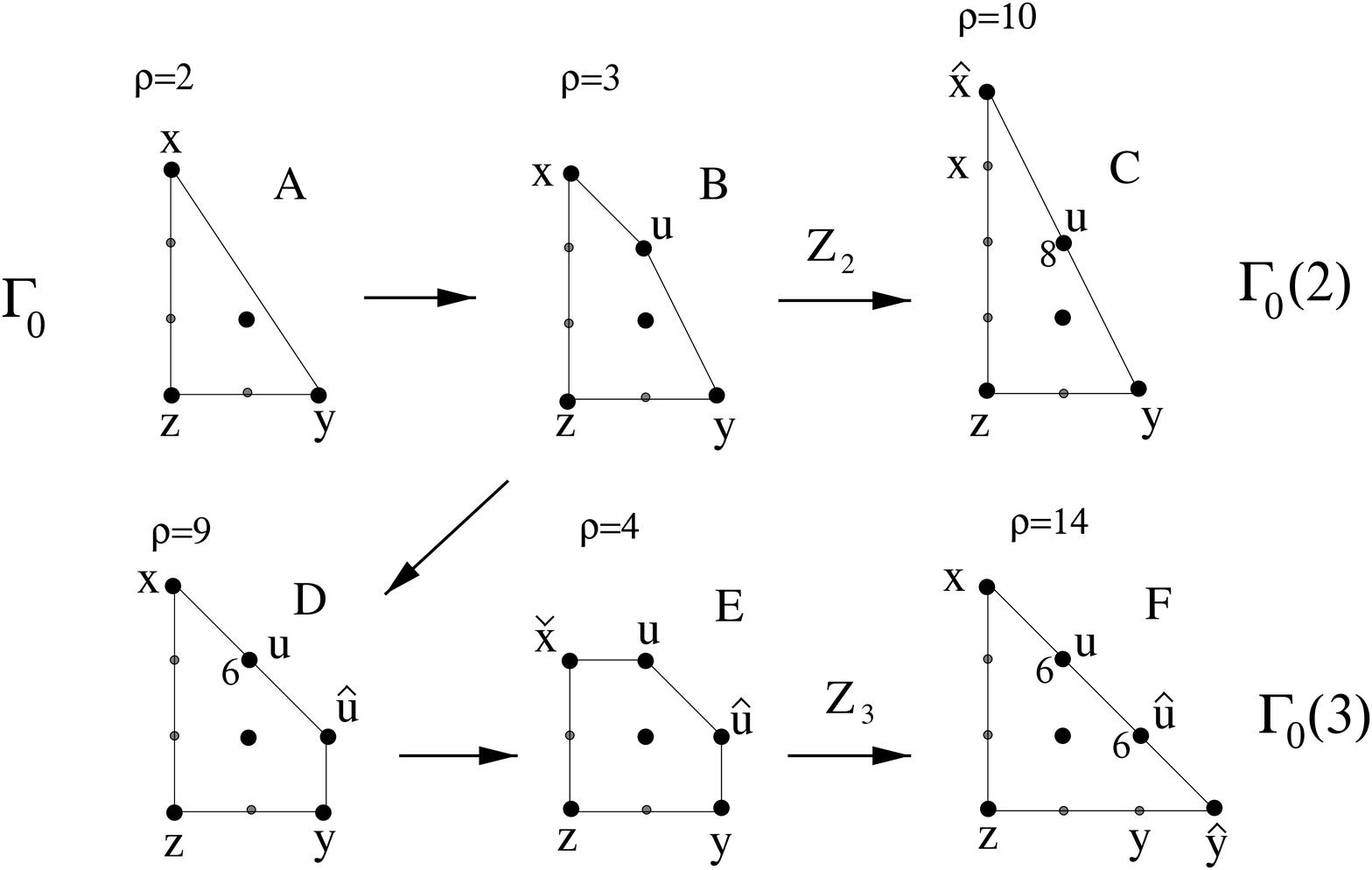}}
\leftskip 1pc\rightskip 1pc \vskip0.3cm
\noindent{\ninepoint  \baselineskip=8pt  
\centerline{{\bf Fig. 1:}
The chains of blow-ups, which lead from a K3 with $\gn$ monodromy to such }
\centerline{with $\Gamma_0(2)$ and $\Gamma_0(3)$ monodromy.
$\rho$=rank(Pic($X$)) and the origin is the interior point.}}
\endinsert}

Adding a point $\nu_u$ to $\{\nu^{\rm coord}\}$, with $\nu_u=a \nu_x+b \nu_y$ means
introducing two new charts $(x_1,y_1)$ and $(x_2, y_2)$ with
$x= x_1^a,y= y_1 x_1^b$ and $x=  x_2 y_2^a, y=y_2^b$
The blow-up~\bup~ correponds to adding the point $u$ with $\nu_u=\nu_x+\nu_y$ in 1.B etc.
$p_B$, $p_C$ are~\batcox~written in the indicated 
coordinates. Spelling out $p_D,p_E,p_F$ can be done analogously. 
The lattice points indicated by small dots are irrelevant as they correspond
to divisors of the ambient space which do not intersect the hypersurface
\foot{Because of $\nu_s,\nu_t$ they lie on a codimension
one face and in the toric construction those divisors do not intersect $X$.}. 
Therefore $x=0$ does not intersect $p_C=0$ and $y=0$ does not intersect $p_F=0$.
The numbers labelling the points indicate how often $x_i=0$ intersects $p$ 
(if different from one). E.g. $u=0$ intersects $p_C=0$ 
in the eight zeros of $f_8$ along a $\IP^1$
and represent the $A_1^8$ read from the descriminant in the previous section. 
Similarly, the sixfold divisors $\{u=0\}\cap \{ p_F=0\}$ and 
$\{\hat u=0\}\cap\{p_F=0\}$ account for the $(A_2)^6$.  
The arrows with $\IZ_2$ and $\IZ_3$ correspond the orbifoldisation, whose
action on the homology of $X^*$ is described in figures 2 and 3. When the 
divisors in the elliptic fiber $\Delta_E^*$  of $X^*$ are contracted, the IIA
theory exhibits the following gauge groups:
$(E_8)^2$, $(E_7)^2$, $(F_4)^2$, $(F_4)^2$,
$(E_6)^2$ and $(G_2)^2$. By blowing up 
$\hat u=y=0$ it is possible to pass from $D$ to $F$. 
However, in the $E$ model the CHL $\IZ_3$ is realised as an outer automorphism
on the cohomology lattice forming $({\hat E}_6)^2$, see figure 3.

\newsec{Gauge groups and heterotic duals}
\subsec{A type IIA perspective}
One way to learn about the possible enhanced
non-abelian gauge symmetries of the above type IIB compactifications
is to analyze the possible fiber degenerations of the
manifolds described there, corresponding to collisions of 
D7 branes as in \mvi,\mvii. Alternatively, we can consider
compactifying on a further $T^2$ to six dimensions 
which is dual to type IIA compactification on the same manifold,
which is in turn dual to the heterotic string on $T^4$.
The higher-dimensional
case can be recovered by considering a limit of the K3 compactification
where the elliptic fiber shrinks to zero size.
In the type IIA compactification 
the gauge symmetry is largely determined by the homology
lattice of 2-cycles. In particular, D2 brane states wrapped on the 
2-cycles give rise to vector multiplets and the intersection lattice
of small 2-cycles coincides generically 
with the Dynkin diagram of the 
enhanced non-abelian gauge symmetry
\cf\ref\kmv{S. Katz, P. Mayr and C. Vafa, Adv. Theor. Math. Phys.
{\ul 1} (1998) 53.}. However, the actual
gauge symmetry can be smaller than what would be expected from 
the lattice of homology cycles, corresponding to the presence
of certain RR $U(1)$ backgrounds on top of the K3 geometry 
\ref\ss{J.H. Schwarz and A. Sen, \plt 357 (1995) 323.}. 
More precisely, the claim of \ss\ is
that if an automorphism of
the homology lattice of K3 is combined with a shift in a 
$U(1)$ factor from the RR sector, the actual gauge symmetry 
is given by the Dynkin diagram of the invariant homology lattice,
corresponding to a folding of the original Dynkin diagram 
spanned by the full homology lattice. We will give further
evidence for this proposal by realizing the appropriate
restricted moduli spaces explicitly in terms of frozen 
parameters in toric geometries.

To see the connection with the higher-dimensional type IIB
backgrounds above, recall that our fibers $\bE i$ support
the $n$-th order shift symmetries $\cx S_i$. These shift symmetries
will identify homology elements of the K3 manifold; thus the
K3's we obtain from fibering symmetric tori with $\IZ_n$ 
shift symmetries will have non-trivial
$\IZ_n$ automorphisms acting on the homology lattice. Finite 
abelian automorphisms of K3 have been classified in 
\ref\nik{V.V. Nikulin, Trans. Moscow Math. Soc. {\ul 2} (19980) 71.}
and used in \cl\ to determine type IIA duals
of CHL heterotic strings in six dimensions. 
   
In order to be able to push these theories up to higher dimensions, the
automorphism has to leave invariant a sub-lattice $\Gamma_{k,k}$ 
which is to be identified with $k$ circles of the heterotic
$T^4$ that we decompactify. Thus, our type IIB theory with 
reduced monodromy group $\Gamma\subset \gn$ should be in one-to-one
correspondence with finite automorphism groups of K3 lattices
preserving a non-trivial $\Gamma_{k,k}$ lattice. In fact, it is not 
difficult to see,
using the results in \nik, that there are  three
automorphisms of this kind which correspond to the three symmetric
tori $\bE i$ given above. In the following we will justify this
identification by a careful study of the singularities of 
the geometries $\bX{2}{i}$.

A geometrical realisation of the $\IZ_2$, which gives rise to the 9d CHL 
compactification, has been given in~\ref\kko{S.\ Kachru, A.\ Klemm 
and Y.\ Oz, {\sl Calabi-Yau Duals for CHL Strings}, hep-th/9712035.} 
and used to study N=1 6d and N=2 4d 
compactifications of the CHL string. These theories can be decompactified 
to 9d.

\subsec{Folding of toric polyhedra}
The appearance of non-simply laced gauge groups
in the type IIB compactification with restricted monodromy 
group has a very clear interpretation when we represent
the Calabi--Yau manifold $X_2$ in terms of toric polyhedra. 

\def\df{\Delta_{E}}
As we have seen at the end of section 2.2, a K3 manifold $X_2$ 
and its mirror $X_2^*$ are described by a 
3 dimensional polyhedron $\Delta$ and a dual polyhedron 
$\Delta^\star$. The vertices $\nu_i$ of $\Delta$ correspond to 
divisors and are 
(taking into account the multiplicity) in one to one correspondence with the 
holomorphic curves in the Picard-Lattice of $X_2$.

Using F-theory/heterotic duality, it was observed in \cf \bv\ and further 
explained in \bv\ that the points above and below the hyperplane $\df$ form the affine
Dynkin diagram of the possible non-abelian gauge symmetry enhancements
from collisions of fiber singularities. 
A microscopic explanation can be given in terms of D2 brane geometries in
type IIA theory \kmv: since points in $\Delta$ correspond to
holomorphic curves, they give rise to states from D2 brane wrappings.
D2 brane states on different 2-cycles $C_1,\ C_2$ will interact if they
have non-vanishing intersection $C_1\cap C_2\neq \emptyset$.
In the K3, mutual intersections of these 2-cycles correspond, 
roughly speaking, to the links (edges) in the toric polyhedron.
Thus, a configuration of points in $\Delta$ forming the Dynkin diagram
of a gauge group $G$ gives rise to a collection of intersecting
D2 brane wrappings. These  generate 
a non-abelian gauge symmetry $G$ in the
space-time theory in the limit where the volume of these 2-cycles 
goes to zero.

\subsubsec{$\Gamma_0(2)$ monodromy}
Let us consider in detail the case with $\Gamma_0(2)$ monodromy
corresponding to the eight-dimensional CHL vacuum with rank reduction
$-8$. We start with the mirror $X^*_B$ of the K3 $X_B$ in sect.~2.2
which describes an ordinary heterotic string compactification
on $T^2$ with $SU(16)\subset SO(32)$ gauge symmetry. Note that going to 
the mirror K3 is necessary to describe the singularity 
structure of $X_B$ in terms of K\"ahler blow-ups corresponding
to holomorphic curves. The polyhedron $\Delta^*_B$ of $X_B^*$ is
shown on the left hand side of Fig.2.\foot{The dark shaded plane 
divides the polyhedron into two identical
halves; only details of the upper half have been displayed.}
 
The shaded triangle
represents the hypersurface $\Delta^*_E$ corresponding to
the elliptic fiber of $X_B^*$. The gauge group $SU(16)$ 
appears via its affine Dynkin diagram $\hat{A}_{15}$ 
composed of the points in $\Delta^*_B$ above the hyperplane $\Delta^*_E$.
There is a second hyperplane $\Delta^*_{\tilde{E}}$ corresponding
to a second elliptic fibration shown in the lower part of Fig.~2.
In this case we read off the gauge group $E_7\times E_7$. 
This of course reflects the fact that in a toroidal 
compactification the $E_8\times E_8$ string lies in the
same moduli space as the $SO(32)$ string\foot{In particular 
these two fibrations are inherited from the two fibrations of
the K3 with maximal singularity $E_8\times E_8$ or $SO(32)$
\ref\canii{P. Candelas and H. Skarke, \plt 413 (1997) 63}.}. 
Both elliptic fibrations are
of the generic type with $\gn$ monodromy.

As explained in sect.~2.2, we can get a K3 $X_C$ with $\Gamma_0(2)$
monodromy by a blow-up of the locus $y=x=0$ in $p_B$. This operation 
corresponds to a $\IZ_2$ modding of the K3. 
The action on the mirror polyhedron $\Delta^*_B$ is shown on the
right hand side of Fig.~2. It acts as a $\IZ_2$ on the homology lattice
which freezes the volumes of pairs of spheres in the $\hat{A}_{15}$
diagram to the same values. Points in $\Delta^*_C$ which correspond
to a pair of spheres are denoted by a black node in Fig.~2. The 
modding folds the Dynkin diagram of $SU(16)$ to that of $Sp(8)$
and similarly that of $E_7\times E_7$ in the second elliptic fibration
to $F_4\times F_4$. The $Sp(8)$ singularity corresponds to the
gauge symmetry of the eight-dimensional CHL vacuum. The $F_4 \times
F_4$ singularity appears in six dimensions upon compactification
of the eight-dimensional CHL vacuum on a $T^2$ with Wilson line \cp.
We now turn to a discussion of how the maximal dimension of a vacuum
can be determined from the geometric data.
\vskip 0.5cm
{\baselineskip=12pt \sl
\goodbreak\midinsert
\centerline{\epsfxsize 5truein\epsfbox{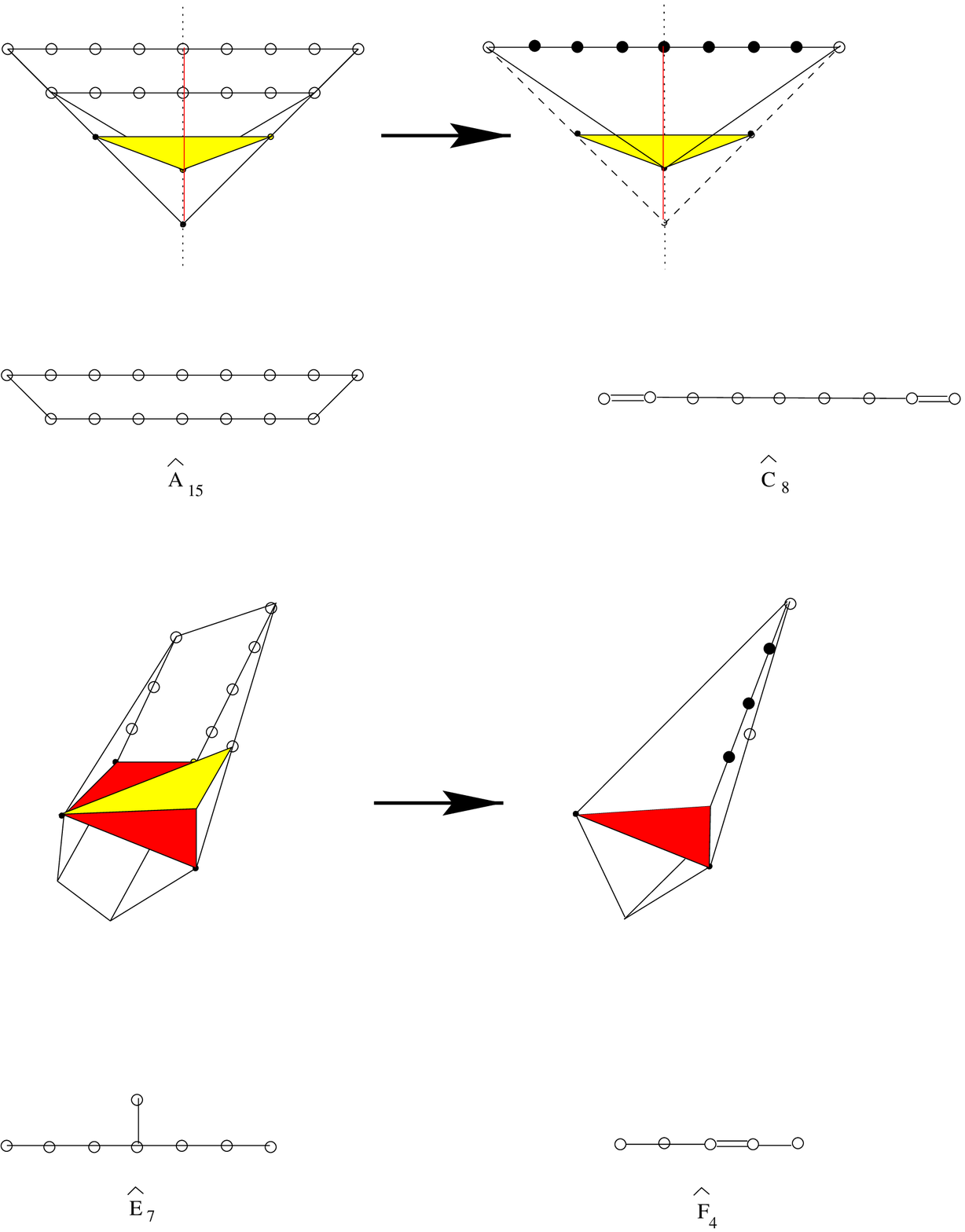}}
\leftskip 1pc\rightskip 1pc \vskip0.3cm
\noindent{\ninepoint  \baselineskip=8pt \centerline{{\bf Fig. 2:}
Folding of K3 polyhedra for F-theory compactification.
}}
\endinsert}
\subsec{Identifying F-theory limits}
To establish the existence of a higher than six-dimensional
vacuum, we have to show that it is possible to find an 
appropriate limit where the fiber of a K3 $\bX{2}{i}$ shrinks
to zero size. This limit corresponds to a decompactification 
of the heterotic string on $T^4$.
{}From the point of view of modding the type IIA theory compactification by
automorphisms of K3 it is clear that the possibility to decompactify
some of the circles of the heterotic dual on $T^4$ implies
the existence of a sublattice $\Gamma_{k,k}$ in the homology of
K3 which is invariant under the automorphism. Moreover, if we
use a direction $\Gamma_{k,k}$ on which the automorphism 
acts as a shift, though we can decompactify in this direction,
the automorphism becomes trivial in the large radius limit and
we land back on a conventional vacuum without a modding. 

As for the first condition, to have an invariant $\Gamma_{k,k}$ 
lattice, it is fulfilled by construction for the elliptically
fibered K3 manifolds with $\Gamma_0(n)$ monodromies that we 
discuss. The question of whether there is a shift action 
on the heterotic $T^2$ that corresponds to the elliptic fiber
can also be read off from the geometric data: if all points
in the hyperplane $\Delta^*_E$ that describes the elliptic
fiber represent a single two sphere in $\bX{2}{i}$, then there is no shift action
on the dual $T^2$ and we can decompactify without interfering
with the action of the modding. If there are points in $\Delta^*_E$
which represent $n$ spheres, there is an $n$-th order shift symmetry 
acting on the dual $T^2$.

\subsubsec{$\Gamma_0(2)$ monodromy: a 9d CHL vacuum}
In the first fibration with $Sp(8)$ gauge symmetry, the 
points in the hyperplane $\Delta_E^*$ correspond to 
single curves in $X^*_C$. We can contract the fiber and
get the dual of the eight-dimensional CHL vacuum. This is
in agreement with the action of the $\IZ_2$ automorphism 
described in \nik,\cl, which leaves invariant a $\Gamma_{4,4}$
lattice with no shifts in a $\Gamma_{3,3}$ part. Thus, we
can actually decompactify to nine dimensions \ref\park{J. Park, \plt
418 (1998) 91.} .

On the other hand, for the second fibration with $F_4\times
F_4$ gauge symmetry we observe that the point representing the
section has multiplicity two corresponding to an order two shift 
on $T^2$. This is in agreement with the fact that in order
to get $F_4\times F_4$ we had to switch on a Wilson line
on a further circle starting from an eight-dimensional CHL compactification
\cp.

\subsubsec{$\Gamma_0(3)$ monodromy: a 7d CHL vacuum}
A polyhedron in the moduli space with $\Gamma_0(3)$ monodromy is shown
in Fig. 3. 
\vskip 0.5cm
{\baselineskip=12pt \sl
\goodbreak\midinsert
\centerline{\epsfxsize 3truein\epsfbox{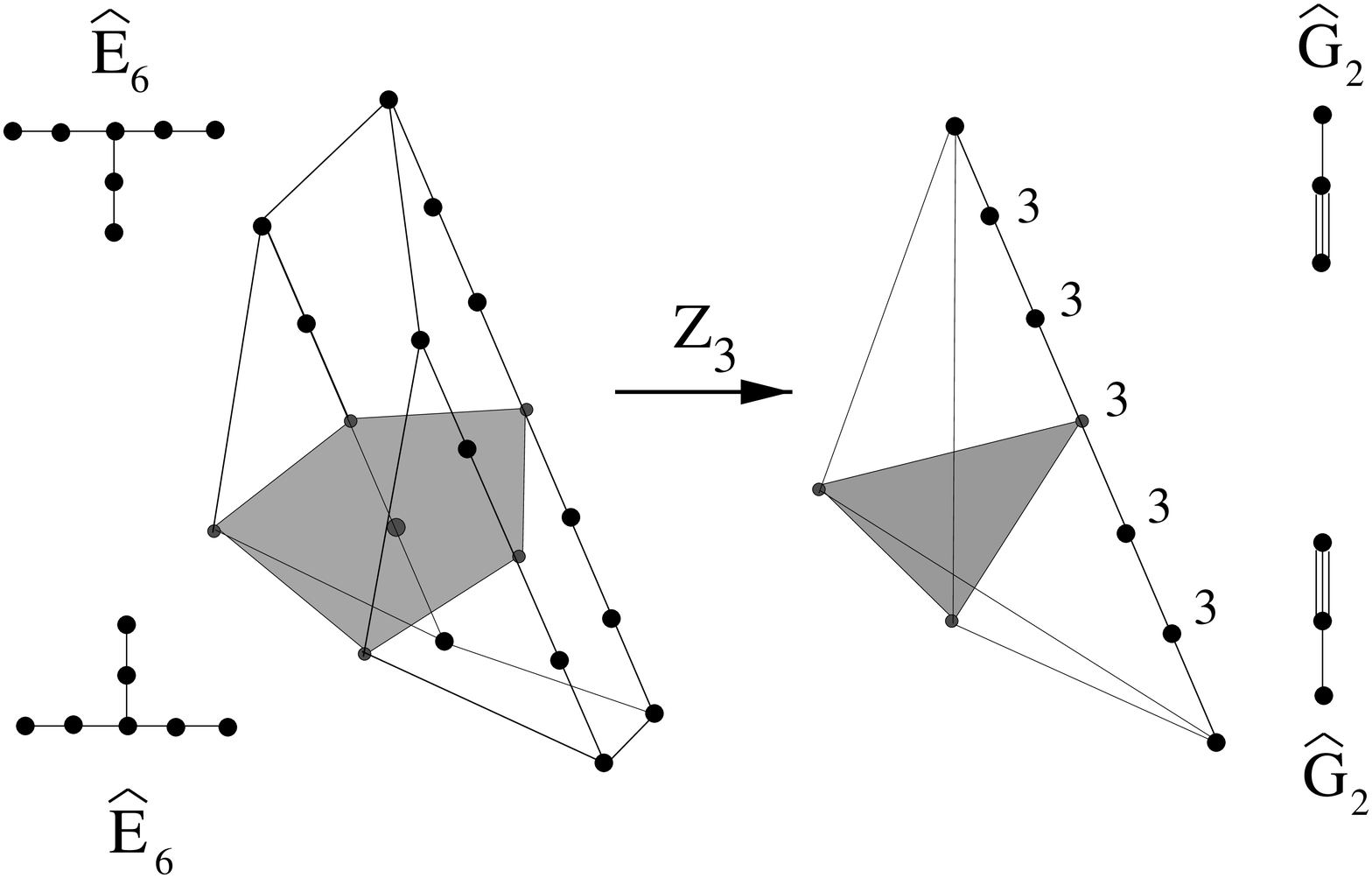}}
\leftskip 1pc\rightskip 1pc \vskip0.3cm
\noindent{\ninepoint  \baselineskip=8pt \centerline{{\bf Fig. 3:}
$Z_3$ folding of the $\hat E_6\times \hat E_6$ K3 to an 
elliptic fibration with $\Gamma_0(3)$ monodromy.}}
\endinsert}
\ni
It describes the dual to a heterotic compactification with
rank reduction $-12$ and $G_2\times G_2$ gauge symmetry.
Note that in this case we have only one elliptic fibration and 
moreover there is a point of order three in the hyperplane $\Delta^*_E$
corresponding to a order three shift on the torus of the heterotic
theory. Therefore we cannot decompactify without destroying the
CHL modding; rather we flow to a point in the moduli space
of ordinary heterotic compactification with $SO(8)^2$ gauge 
symmetry in this limit. This is in agreement with the $\IZ_3$ action
described by Nikulin, which leaves invariant a $\Gamma_{2,2}$ lattice
but acts as a shift on a $\Gamma_{1,1}$ part of it.

\subsubsec{$\Gamma_0(4)$ monodromy}

Since the $\IZ_4$ action leaves a $\Gamma_{3,3}$ torus invariant
but acts as a shift on a $\Gamma_{1,1}$ sublattice, we are lead
to conclude, in analogy  with the cases considered above, that we can
decompactify to eight dimensions. However due to the fact that the elliptic fiber 
is represented as a complete intersection, the description of the 
structure of the sections and the singularities on them is not 
straightforward. Hence it is hard to distinguish whether this is a new 8d 
theory or whether it merely flows in the F-theory limit 
to a special point in the moduli space of an already known one.

\subsec{Frozen moduli spaces in toric geometry}
We have seen that the toric geometries reproduce quite well
the moduli spaces of reduced rank of the heterotic theory.
We will give now a more precise formulation of this fact 
using some more technology of toric geometry. In fact, there
is a well known phenomenon in Calabi--Yau manifolds constructed
as hypersurfaces in toric varieties which has been considered
mostly as a technical subtlety so far: the appearance of so-called
non-toric divisors and non-polynomial deformations. This
situation refers to the fact that a given toric representation of
a Calabi--Yau manifold $X$ might not allow for a general deformation
but some of the K\"ahler or complex structure moduli are frozen 
to a specific value. What we suggest is that rather than being a
technical subtlety, these restricted Calabi--Yau moduli spaces
are in one-to-one correspondence with physical moduli spaces,
where some moduli are stuck due to background fields. As was described
in section~3.1 we obtain a K3 
with $\Gamma_0(n)$ monodromy from a ${\bf Z}_n$ modding. This modding
removes points from the dual polytope
$\Delta^\star$ of $X$, reducing the rank of the gauge group.
In $\Delta$ this has the effect of introducing so called non-toric
deformations. Although the K3 still has a total of 20 K\"ahler and
complex structure deformations, a
certain number of them cannot be
represented toricly. That is from a toric description the sizes of 
two-cycles are set to identical 
values\foot{An example of this phenomenon is
that of type IIA compactified on the elliptically fibered Calabi-Yau
three-fold with base $F_2$ in which there is one non-toric deformation. 
This accounts for the strongly coupled $SU(2)$ with a massless
adjoint.}.
It is this frozen F-theory geometry which accounts 
for the reduction of rank leading to the CHL dual.

In the conventional heterotic case, with vector structure, 
the total number of K3 deformations from both, K\"ahler
and complex deformations, adds up to $dim\ H_{1,1}(K3)=20$
corresponding to the Narain lattice $\Gamma_{18,2}$. 
Reducing the rank of the gauge group in the fibration 
by reducing the number of points of a dual polyhedron
$\Delta^\star$, or equivalently the number of K\"ahler
deformations, is compensated by the correct number of
new complex deformations corresponding to new
points in the polyhedron $\Delta$ describing the mirror K3.
However, in the present case, the 
reduction of K\"ahler moduli is {\it not} compensated by
the complex deformations leading to a reduced number of K3
moduli. 

Let us review briefly the appearance of non-toric and non-polynomial
deformations.\def\Xd#1{X_{#1}(\Delta)}
An $n$-dimensional toric hypersurface $\Xd n$ is  
defined as the zero locus of the 
section of an appropriate line bundle on a non-singular 
toric variety $X(\Sigma)$. The $n+1$ dimensional toric 
variety $X(\Sigma)$ is defined as the quotient of 
$\IC^N\backslash F$ by $(\IC^\star)^{N-n-1}$. If $x_i$ are coordinates
on $\IC^N$, the disallowed set $F$ is given by subsets
$\bigcap_{k} x_{i_k}=0$ related to fixed points of the 
scaling actions. The simplest example is $\IP^N$ with
$N+1$ coordinates $x_i$, one $\IC^\star$ action $x_i \to \lambda x_i$
with the fixed point $x_i=0\ \forall i$ omitted.

\def\Xds#1{X^\star_{#1}(\Delta^\star)}
The toric divisors $D_i:x_i=0$ generate a basis for 
$H^n(X(\Sigma))$. Given a representation $\Xd n$
of a Calabi--Yau manifold $X_n$
as a hypersurface in $X(\Sigma)$, the toric divisors
$D_i$ descend to elements $D^X_i$ in $H^{n-1}(\Xd n)$ from the intersection
$D_i\cap X$. However, if it happens that the intersection
$D_i\cap \Xd n$ contains $l$ disconnected components $D^{X,\alpha}_i,\ 
\alpha=1,\ldots,l$
in $X_n$, a priori independent elements of $H^{n-1}(X_n)$ 
are represented by the same class $D_i$ in the ambient space.
As a consequence, the a priori different volumes of the 
divisors $D^{X,\alpha}_i$ are frozen to the same value.
Thus, the hypersurface $\Xd n$ can only represent a restricted 
subset of the Calabi--Yau moduli space $X_n$. 

To a hypersurface $\Xd n$ defined as a hypersurface in a toric
variety corresponds a polyhedron $\Delta$ of lattice points
in a standard lattice $M\sim\IZ^{n+1}$; c.f. the discussion in sect.~2.2. 
In particular, the polyhedron $\Delta$
encodes in a very efficient way information about the cohomology
and the intersection properties of $\Xd n$. Moreover, there
is a canonical construction of the mirror manifold $\Xds n$ of
$\Xd n$ in terms of the dual polyhedron $\Delta^\star$.
In 
\ref\bat{V. Batyrev, Journal Alg. Geom. {\ul 3} (1994),
Duke Math. Journ. {\ul {69}} (1993) 349} 
Batyrev shows 
that the Hodge numbers $h^{1,1}$ are determined by the polyhedron via
\def\del{\Delta}\def\dels{\Delta^*}
{\ninepoint{
\eqn\hodgenum{{\eqalign{
h^{1,1}(\Xd n)&=h^{n-1,1}(\Xds n)=
l(\dels)-(n+2)-\sum_{{\rm codim} S^\star=1}l^\prime(S^\star)+
\sum_{{\rm codim} S^\star=2}l^\prime(S^\star)\cdot l^\prime(S)\ }}}}}
%

\noindent
and similarly for $h^{n-1,1}(\Xd n)=h^{1,1}(\Xds n)$ with the
roles of $\del$ and $\dels$ exchanged.
Here $S$ denotes faces of $\Delta$ and $S^\star$ the dual face of 
$S$. $l$ and $l^\prime$ are the numbers of integral points on a 
face and in the interior of a face, respectively. 
The last term is a correction term corresponding to 
the non-toric deformations. We will be mainly interested
in the cases $n=2,3$. For K3, $h^{1,1}$ gives the rank of
the Picard lattice of $\Xd 2$ and a non-zero correction corresponds
to having simultaneously points on a one-dimensional 
edge $S$ of the 3-dimensional
polyhedron $\del$ and on the dual edge  $S^*$ in $\dels$.
In the 3-fold case, a non-zero correction term arises
from having simultaneously points on a face(edge) $S$ of
$\del$ and the dual edge(face) $S^*$ of $\dels$. 

In general, if we freeze the
volumes of two, say, two-cycles to the same value, nothing 
interesting happens. However, the existence of a hypersurface
$X^\prime_n(\Delta^\prime)$ 
with restricted moduli space corresponds to the fact
that apart from the volumes, {\it these two-cycles share identical 
intersection properties}. In this case we are at a point with
extra symmetry and $X^\prime_n(\Delta^\prime)$ corresponds to the modding 
of this symmetry. Note that in general we do not expect the 
special locus in the moduli space to correspond to a singular
manifold, but rather to a particularly symmetric
configuration. This is analoguous to the fact that in the heterotic
theory the CHL construction requires the presence of 
several gauge group factors broken in an identical manner,
but not a restoration of non-abelian gauge symmetries.

\subsec{Heterotic compactifications to six dimensions with $U(1)$ backgrounds}
So far we have interpreted the restricted K3 moduli spaces in terms
of type IIA compactifications. However, we could also ask about their
meaning if we choose to compactify a heterotic string to  get an 
$N=1$ six-dimensional vacuum. Anomaly free $N=1$ theories in six 
dimensions have been described in 
\ref\gsw{M.B. Green, J.H.Schwarz, \nup 254 (1985) 327.}.
A perturbative
vacuum requires a combination of abelian and
non-abelian gauge instantons on K3 of total instanton
number 24. In addition, minimization of the action of
instantons on K3 requires 
\ref\witsu{E. Witten, \nup 268 (1986) 79.} 
\def\bx#1{\bar{#1}}
\eqn\uyi{
F_{ij}=F_{\bx  i \bx  j}=g^{i\bx  j}F_{\bx  i j }=0\ ,}
implying 
\eqn\uyii{
\int_{X_2} J\wedge {\rm tr }\ F =0.}
The latter equation implies a non-trivial interplay of
abelian instantons and K3 moduli; in particular, decomposing
the $U(1)$ flux in harmonic $(1,1)$ forms $\omega^\mu$ on $X_2$, $F^{(a)}=
\alpha^{(a)}_\mu \omega^\mu$, and similarily for the
Kaehler form, $J=t_\mu \omega^\mu$, eq. \uyii\ gives a linear
constraint on the K\"ahler moduli
\eqn\cons{
\alpha^{(a)}_\mu t_\nu C^{\mu \nu}=0 \ ,
}
where $C^{\mu \nu}$ is the intersection form on $X_2$. This
constraint leads to vacua with frozen K\"ahler moduli. Note 
that we cannot simply restrict to a frozen configuration
in a generic K3 moduli space, but we should provide
a consistent special geometry based only on  the remaining 
perturbations. Moreover, the freezing should involve full
vector multiplets, rather than only their scalar components.
By duality, this freezing should correspond again to F-theory 
compactification on a Calabi--Yau manifold with restricted
moduli space in {\it hypermultipets}. Therefore we arrive
at the picture that the heterotic string on a K3 $X_2$ with frozen
moduli due to $U(1)$ instanton background is dual to 
F-theory on $X_3$ with non-polynomial deformations.

Note that from the F-theory point of view,
CHL vacua and six-dimensional $U(1)$ instantons 
are on a very similar footing due to the fact
that both bundle moduli and geometric moduli of the
heterotic string get mapped to geometric data of the
F-theory compactification.

Type IIA compactification on Calabi--Yau three-folds with special 
fiber models for the elliptic fiber has been considered in 
\ref\kv{S. Kachru and C. Vafa, \nup 450 (1995) 69.}%
\ref\kmvii{A. Klemm, P. Mayr and C. Vafa,
{\it BPS states of exceptional non-critical strings},
to appear in the proceedings of the conference {\it Advanced
Quantum Field Theory}, (in memory of Claude Itzykson),
CERN-TH-96-184, hep-th/9607139.}%
\ref\ald{G. Aldazabal, A. Font, L. Ibanez and A.M. Uranga, \nup
492 (1997) 119}%
\ref\caniii{P. Candelas, E. Perevalov and G. Rajesh,
\nup 502 (1997) 594.}.
However, these models can generically not be pushed to a six-dimensional F-theory 
compactification dual to heterotic theory with $U(1)$ backgrounds
due to the generic monodromy of the elliptic fibration\foot{Our 
conclusion concerning this point is different from the one suggested in \ald .}.  
Most of these models correspond rather to five-dimensional 
compactifications with $U(1)$ Wilson lines on an extra circle and therefore 
flow to the moduli space of heterotic 
compactifications without $U(1)$ backgrounds in six dimensions.
In fact this is clear from the presence of the corresponding transitions
between Calabi--Yau manifolds \kmvii.
Note that analysis of the four-dimensional supergravity spectrum
cannot tell the difference between a five-dimensional 
compactification with $U(1)$ Wilson lines and a six-dimensional
compactification with $U(1)$ instantons. The difference is
precisely detected by the moduli frozen by the constraint \uyii\ 
and severely restricts the possible vacua of this type.

\subsec{Enriques involution and CHL orbifold}

The F-theory dual of a rank $r$ CHL compactification 
involves naturally a restricted family of elliptic $K3$ surfaces $X$, 
whose Picard lattice has rank $\rho=20-r$. 
The complex structure moduli space will then be of the general form of a locally symmetric 
space~\ref\bpv{W.\ Barth, C.\ Peters and A. Van de Ven, {\sl Compact complex Surfaces}, 
Springer Berlin (1984).}  
\eqn\algmod{{\cal M}=\Gamma_{T}\backslash O(2,20-\rho)/O(2)\otimes O(20-\rho),}  
identified by discrete automophisms $\Gamma_{T}$ of the transcendental lattice 
$T$, which has signature $(2,20-\rho)$. 
In particular, for the $\IZ_2$ CHL construction in $8d$, 
in the following just denoted by CHL, 
we have argued that the required $\Gamma_0(2)$ monodromy restricts the dimension of the 
complex structure moduli space to ten and found, besides the 
hyperbolic factor $H$ required by the fibration 
structure, further eight algebraic curves in the $A_1^8$. 

There is a well known example of $K3$ surfaces whose Picard group has rank ten,
namely the double cover of an Enriques surface. Every Enriques 
surface $Y$ has a $K3$ $X$ as an unbranched double cover and can thus be given 
as $Y=X/\sigma$ with  $\sigma$ fixed-point free. The $(2,0)$ form is anti invariant
under $\sigma$ and therefore  $c_1(Y)\neq 0$ is a torsion class: $c_1^2(Y)=0$. 
$Y$ has no holomorhic periods, but $X$ can be used to define its period map and local 
and global Torelli theorems have been proven from this for
$Y$~\ref\horikawa{E.  
Horikawa, {\sl On the Periods of Enriques Surfaces I}, Math. 
Ann. 234, 73-88 (1978).}, see also \bpv. 
One can identify the induced action of the covering involution 
$\sigma$ on the cohomology lattice 
\eqn\lambda{\Lambda(X)=(-E_8)\oplus (-E_8)\oplus H \oplus H \oplus H,}
as $\rho(x\oplus y\oplus z_1\oplus z_2 \oplus z_3)=
(y\oplus x \oplus -z_1\oplus z_3 \oplus z_2)$ after choosing 
a compatible marking $\phi:H^2(X,\IZ)\rightarrow \Lambda$, i.e. with
$\phi \circ \sigma^*=\rho \circ \phi$. $H$ is the hyperbolic 
lattice spanned by $e_1$, $e_2$ with $e_1\cdot e_2=1$ $e_1=e_2^2=0$ and 
$E_8$ denotes the $E_8$ lattice. Under $\rho$, $\Lambda$ splits
\eqn\split{\Lambda=\Lambda^+\oplus \Lambda^-} 
into an invariant part $\Lambda^+=(-2 E_8)\oplus 2 H$, 
which comes from the Picard lattice $S$ of $X$ and the noninvariant 
part $\Lambda^-=(- 2 E_8)\oplus 2 H\oplus H$, which comes from the
transcendental lattice $T$ in the cohomology of $X$. 
The cohomology lattice of $Y$ is isometric to  
$\Lambda^0=\Lambda^+/2=(-E_8)\oplus H$. Since $\omega \in \Lambda^-$  
and the period integral $([\omega], \gamma)=\int_\gamma \omega$ 
is invariant under $\sigma$ it must vanish for $\gamma$ Poincare dual 
to an element of $\phi^{-1}(\Lambda^+)$. The period points, describing
the complex moduli space   
of $X$ all belong to $\Omega^-=\{[\omega]\in \IP(\Lambda^-\otimes \IC)| 
(\omega,\omega)=0, (\omega,\bar \omega)>0\}$, which contains two
disjoint copies  
of a 10d bounded domain as in~\algmod\foot{In fact for $Y$ not all 
of~\algmod~is mapped out by period points. It is known that precisely 
the hyperplanes $H_d=\{[\omega]\in \Omega^-|(w,d)=0\}$ with $(d,d)=-2$ have to be 
subtracted~\horikawa\bpv. }, where the discrete group $\Gamma_{\Lambda^-}$ 
is the restriction to $\Lambda^-$ of those elements $g$ in $Aut(\Lambda)$ which are 
compatible with \split, i.e. $g\circ \rho=\rho\circ g$. This is
very close to the description of the moduli space for the CHL in \cp, 
namely~\algmod~with $\Lambda^-$ replaced by the invariant part of the Narain 
lattice $\Lambda^{\rm CHL}=(-2 E_8)\oplus H\oplus H$ under the CHL involution. 
In fact the difference is the scale 2 on one of the hyperbolic factors due
to the $(z_2,z_3)$ exchange in $\rho$, which has no analog in  the CHL modding. 
However Nikulin describes in \ref\nikulinI{V.\ V.\ Nikulin, {\sl On the Quotient groups
of the automorphism groups of hyperbolic forms modulo subgroups generated by 2-reflections},
J.\ Sov. Math {\bf 22} (1983) 1401} a closely related $Z_2$ operation $\sigma_N$ 
under which the $(2,0)$ form is like wise anti-invariant but, as $(z_2,z_3)$ is not 
exchanged, has now indeed $T=\Lambda_N^-=  (-2 E_8)\oplus H\oplus H$, while 
$S=\Lambda^+_N=(-2 E_8)\oplus H$. $\sigma_N$ is not fixpoint free, but fixes 
two elliptic fibres. 

Horikawa \horikawa~gives a construction of $X$ as double covering over 
$F_0=\IP^1\times \IP^1$ branched at $B$, a bidegree $(4,4)$ hypersurface. 
The generic $B$ has $25$ perturbations, seven reparametrisations have to be 
substracted, which yields $18$ complex structure deformations for $X$. 
The $\IZ_2$ involution $\hat \sigma$ acts on the coordinates of $F_0$ as 
$\hat \sigma: (x:z,u:v)\mapsto (x:-z,u:-v)$. 
The involution $\sigma$ is the combination of $\hat \sigma$ with the
covering involution $y\mapsto -y$. We have $13$ $\sigma $ invariant monomials: 
$x^{4-2i} z^{2i} u^{4-2j} v^{2j}$ $i,j=0,1,2$ and
$x^{3-i} z^{1+i}u^{3-j} v^{1+j}$ $i,j=0,2$. The $SL(2,\IC)^2$ 
of the two $\IP^1$'s is broken to two parameters and the overall 
scaling removes an other parameter. Hence we have ten complex structure 
deformations. This $X$ has a nice toric description by the polyhedra $(\Delta,\Delta^*)$, 
which are shown in figure 4. X admits at least two and $X^*$ at least eight 
different elliptic fibrations, as inspection of the toric diagram  shows. 
We pick the one in which the torus is represented by $y$ as the double cover of a 
quartic in $x,z$ and in which $u,v$ become the base coordinates.
Now we mod out the $\IZ_2^{\rm CHL}$, which was described in~\kko, namely the 
elliptic involution $y\mapsto -y$, combined with inversion of the 
inhomogeneous $\IP^1$ coordinate $w\mapsto {1\over w}$, which acts in 
diagonalized form as $v\mapsto -v$. On the other hand $\sigma_N$ is identified with 
$\sigma_N: (x:z,u:v)\mapsto (x:z,u:-v)$ with the two fixed fibers located over
$(u,v)=(1,0),(0,1)$. The elliptic involution that we do in addition in 
$Z_2^{CHL}$ to keep the holomorphic form, exchanges $\Lambda^+_N$ with 
$\Lambda^-_N$, hence what we keep is $\Lambda^-_N=\Lambda^{CHL}$. 
The relation between the K3 $X_N$ admitting $\sigma_N$ as involution to the double cover 
of Enriques has been pointed out in \ref\grnik{V.\ A. Gritsenko and V.\ V.\ Nikulin,{\sl K3 Surfaces, 
Lorentzian Kac-Moody Algebras and Mirror Symmetry},  alg-geom/951008.} (see sect. 6, example 3). 
The mirror operation identifies $T(X^*)=2H\oplus S(X)$ and $S(X^*)=T(X^*)^\perp=2H\oplus( -2 E_8)$ 
hence it is $X_N^*$ that is related to the double covering of an Enriques surface.

The orbifold $X/Z_2^{CHL}$, a rank(Pic)=10 K3, can be readily defined by the reflexive pair 
$(\Delta',\Delta^{*\prime})$, which is obtained, as in~\kko , from $\Delta^*$ by keeping only solid points, 
which correspond to the invariant monomials, see Figure 4. This orbifold has two $A_1^4$ 
singularities on two different sections, as can be seen from the points of multiplicity 
$4$ in $\Delta'$. Also note that $X^*$ has an $Sp(4)^2$ gauge symmetry.

\vskip 0.5cm
{\baselineskip=12pt \sl
\goodbreak\midinsert
\centerline{\epsfxsize 3.5truein\epsfbox{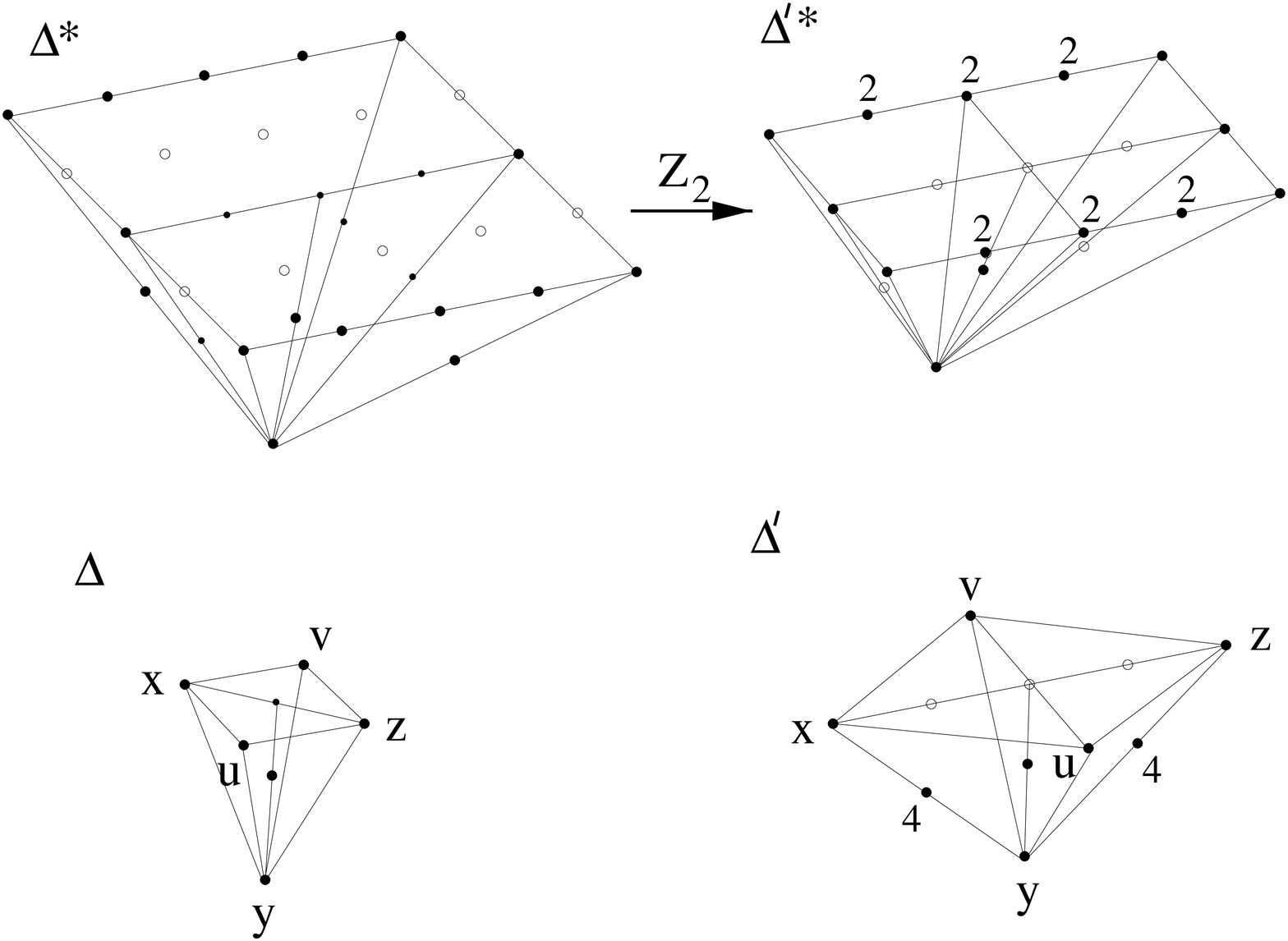}}
\leftskip 1pc\rightskip 1pc \vskip0.3cm
\noindent{\ninepoint  \baselineskip=8pt  
\centerline{{\bf Fig. 4:}
$\IZ_2$ modding of a rank(Pic)=18 elliptic fibration K3 to the rank(Pic)=10 CHL dual.}}
\endinsert}

Finally note that by a series of blow-ups and blow-downs, all within the class
of rank(10) $\Gamma_0(2)$ K3's, we can recover the $Sp(8)$ case discussed 
in section 3.2. The corresponding blow ups take place by adding and removing
points in the $(xzuv)$-plane of $\Delta'$. Below we indicate the gauge groups
which occur in this process.  Fixed are the points $x,z$ and the configuration
$(xzuv)$ is the starting configuration with $Sp(4)^2$ from figure 4. One gets 
$(u\hat u v)$: $Sp(3) \times Sp(4)$, $(u \hat u v\hat v)$: $Sp(2) \times Sp(4)$, 
$(\hat u v \hat v)$: $Sp(2) \times Sp(5)$,
$(\hat u \hat v)$: $Sp(2)\times Sp(6)$, $(\hat u  \hat v {\hat {\hat v}})$ $Sp(6)$, 
$( \hat u \hat {\hat u} \hat {\hat v})$ $Sp(7)$ and $(\hat{\hat u},\hat{\hat v})$ 
gives the $Sp(8)$ we discussed in  section 3.2.

\vskip 0.5cm
{\baselineskip=12pt \sl
\goodbreak\midinsert
\centerline{\epsfxsize 1.5truein\epsfbox{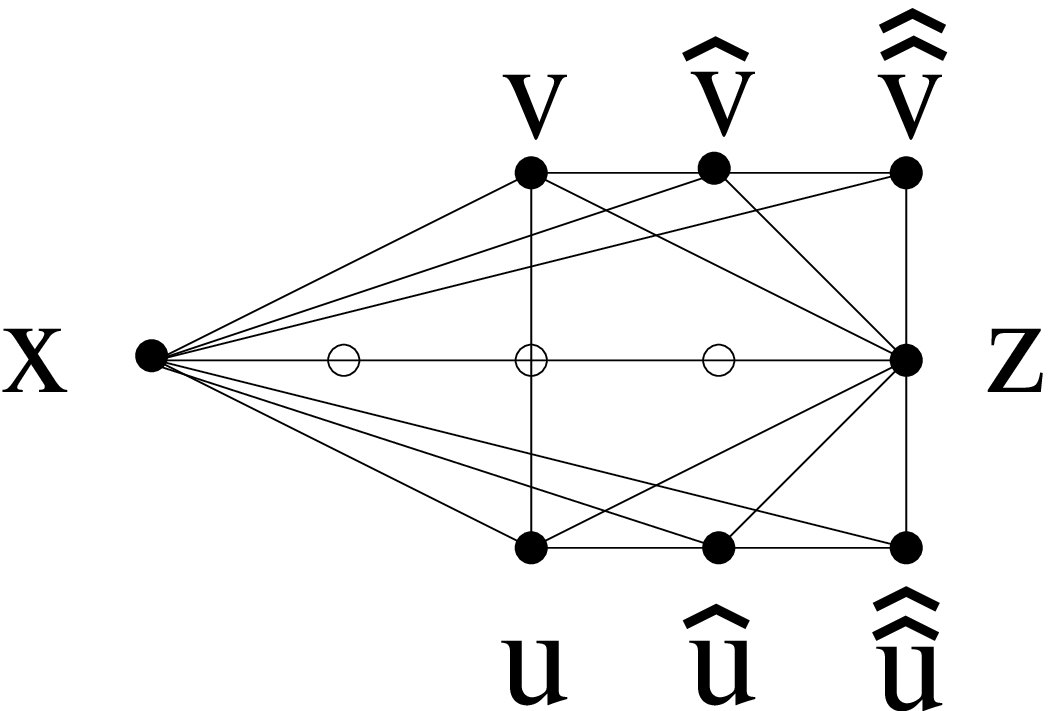}}
\leftskip 1pc\rightskip 1pc \vskip0.3cm
\noindent{\ninepoint  \baselineskip=8pt  
\centerline{{\bf Fig. 5:}
Elementary transformations within the $\Gamma_0(2)$ K3.}}
\endinsert}

\newsec{Moduli space of $G$ bundles without vector structure on elliptic curves}
\def\tw{\tilde{w}_2}
Having identified F-theory geometries corresponding to heterotic CHL models
in eight dimensions, leads a very nice application for the description of
the moduli space of $G$ bundles over elliptic curves. In 
\witwov\ it was
shown that CHL string constructions in eight dimensions 
can be interpreted as heterotic compactifications on $T^2$ with a vector
bundle without vector structure. E.g. for the heterotic theory with 
gauge group $G=Spin(32)/\IZ_2$ in ten dimensions, it is consistent
to have backgrounds obstructing the presence of the vector representation 
of $SO(32)$,  due to the absence of perturbative states in this representation.
The non-triviality of the action on the vector representation is measured by
a mod two cohomology class $\tw$ which integrates to -1 on a two-cycle 
on which vector structure is obstructed. 

We are interested in the moduli space of such bundles over the elliptic curve
$E$ on which the heterotic string is compactified. These moduli spaces
turn out to be described by weighted projective spaces 
\ref\loj{E. Looijenga, Invent. Math. {\ul 38} (1977) 17;
Invent. Math. {\ul 61} (1980) 1.}
and have been analyzed in detail in the mathematical literature 
\ref\fmw{R. Friedman, J. Morgan and E. Witten, \cmp 187 (1997) 679.}.
In \kmv\ 
it was realized that one can get a very simple and canonical description of 
these moduli spaces in terms of complex geometries as a consequence of 
heterotic/type 
IIA duality together with mirror symmetry. A simple reasoning 
goes as follows: F-theory on elliptically fibered K3 with a singularity $H$
from collisions of singular fibers is dual to heterotic string on $E$ with
a vector bundle taking values in $G$, the commutant of $H$ in the primordial
gauge group $E_8\times E_8$ or $SO(32)$. On compactification on a further $T^2$
to six dimensions, we blow up the $H$ singularity in K\"ahler moduli and
obtain type IIA on the smooth K3 $X(H)$, dual to heterotic theory on $T^2\times 
T^2$,
where now we have in addition to the $G$ Wilson lines on the first $T^2$ also
$H$ Wilson lines on the second torus, breaking to the generic abelian gauge
group $U(1)^{24}$. Thus, the moduli space of K\"ahler blow-ups of the $H$ singularity
in the K3 $X(H)$ is isomorphic to the moduli space of $H$ Wilson lines on the 
second $T^2$.
Application of mirror symmetry gives type IIA on the mirror K3 $X(G)$, with the 
roles
of complex and K\"ahler moduli exchanged. Thus, the complex deformations of 
$X(G)$
give a realization of the moduli space of $H$  bundles on $T^2$.

Let us apply this procedure to the K3 dual to heterotic CHL string
in eight dimensions. The only difference
compared to \kmv\ is that we have to apply mirror symmetry to the 
local $Sp(8)$ singularity with symmetric fiber. After a change of variables,
the defining equation for $X$ can be rewritten as 
\eqn\pee{
p=a_0v^{-1}+(y^2+x^4+z^4+\alpha yxz)+v(z^8+z^7x+\ldots+x^8)=0\ .}
To concentrate on a neighborhood of
the local singularity mirror to the $Sp(8)$ singularity
in the elliptic fiber of the mirror K3, we have to set the coefficient $a_0$
of the $v^{-1}$ term to zero. In this limit, we can integrate out $v$ 
and obtain the spectral cover description of \fmw:
\eqn\peee{
p_E=y^2+x^4+z^4+\alpha yxz=0,\qquad p_C= z^8+z^7x+\ldots+x^8=0\ .}
These equations describe eight points given by the zeros of $p_C$ on the
elliptic curve defined by $p_E=0$. Actually, since there is no $y$ dependence
in $p_C$, we have to divide by the $\IZ_2$ operation $y\to -y$ and get 
eight points on the orbifold $T^2/\IZ_2$. This is the usual description 
of $Sp(8)$ bundles on an elliptic curve $E$ \fmw. However, since
the torus $p_E$ is not of the generic form,  we have in addition to take 
into account the shift symmetries \shiftone; this gives further identifications 
on $T^2/\IZ_2$ by the two half-period shifts $z\sim z+\fc{1}{2},\ 
z\sim z+\fc{\tau}{2}$ in a lattice basis with periods $(1,\tau)$. Therefore
the points $p_C=0$ are defined on a orbifold $T^2/\IZ_2$ with the
torus lattice rescaled by a factor two in each direction. This agrees 
perfectly with the description in ref. \witwov, where it has been argued
via different reasoning that the dual orientifold lives on a half-sized 
torus. This gives a further independent check on our identification
of the F-theory geometry.

The above description can be easily generalized to arbitrary $Sp(n)$ 
bundles using the local mirror constructions of \kmv\ with a symmetric
torus $\bE2$ as the generic elliptic fiber of the elliptic singularity.
It would be interesting to work out a more complete description for 
other gauge groups as well as for the other tori $\bE3$ and $\bE4$.

\ni{\bf Acknowledgments}\br
We would like to thank W. Lerche, D. Lowe, D.\ Morrison, 
J. Polchinski, H.\ Skarke, S. Stieberger and  
A.\ Todorov for valuable discussions. The work of P.B. and P.M. was supported in part by
the Natural Science Foundation under Grant No. PHY94-07194.
S.T. is supported by GIF, the German-Israeli Foundation for Scientific 
Research and by the European Commission TMR programme
ERBFMRX-CT96-0045. A.K., P.M. and S.T. thank the ITP in Santa Barbara for
hospitality. 

\vfill\eject

\noindent
{\bf Appendix A: Congruence subgroups $\Gamma_0(n)$ and $\Gamma(n)$}\br
Let us collect some relevant properties of the subgroups
$\Gamma_0(n)$ and $\Gamma(n)\subset\Gamma_0(n)$ of $\gn=SL(2,\IZ)$.
More details can be found in 
\ref\kob{N. Koblitz, {\it 
Introduction to Elliptic Curves and Modular Forms}, (Springer Verlag, 1984).}.
The definitions of $\Gamma_0(n)$ and $\Gamma(n)$ are
\eqn\defsg{
\eqalign{
\Gamma_0(n)&=\{ \pmatrix{a&b\cr c&d\cr} \in \gn,\ 
c=0\ {\rm mod}\  n \}\cr
\Gamma(n)&=\{ \pmatrix{a&b\cr c&d\cr} \in \gn,\ 
a=d=1\ {\rm mod}\ n,\ b=c=0\ {\rm mod}\  n \}\cr
}}
The corresponding inhomogeneous groups are denoted by 
$\bar\Gamma_0,\,\bar\Gamma_0(n)$ and $\bar\Gamma(n)$.
{}From the definition \defsg\ we see that $\Gamma_0(n)$ fixes
the $n$ order $n$ points $(k/n,0)\ {\rm mod}\ (r,s),\ k=0,\dots,n-1,\ 
r,s \in \IZ$, while $\Gamma(n)$ fixes the $n^2$ points 
$(k/n,l/n)\ {\rm mod}\ (r,s),\ k,l=0,\dots,n-1,\ r,s \in \IZ$. For
the elliptic curve these coordinates refer to the basis spanned 
by the periods $(\omega_1,\omega_2)$ with $\omega_i=\int_{\alpha_i}\omega$. 

A special case appears for $n=0\ {\rm mod}\ 2$, since in this case
the points of order two are preserved, which are the branch 
points of the double covering of the Riemann sphere. 
By $\wp({\omega_i\over 2})=e_i$ with  
$\omega_i=\omega_1,\omega_2,\omega_1+\omega_2$ these points
are identified with the roots of the Weierstrass form  
$(\wp')^2=4\prod_{i=1}^3(\wp-e_i)$  
for $E$ and each invariant root gives rise to a global section of the
elliptic fibration. Therefore for $\Gamma_0(2n)$ we have 
two global sections, the infinity section $y=\wp'=\infty,\ 
x=\wp=\infty$ corresponding 
to $(0,0)$ as well as a further global section $(\fc{1}{2},0)$,
whereas for $\Gamma(2n)$ we have four global sections 
corresponding to the four branch points $(0,0),\ (\fc{1}{2},0),\ 
(0,\fc{1}{2}),\ (\fc{1}{2},\fc{1}{2})$.

\vskip.5cm
\noindent
{\bf Appendix B: Restricted monodromies}\br
In this appendix we summarize degenerations and  monodromies 
which occur for the fibrations $\bX{2}{2}$, $\bX{2}{3}$ and $\bX{2}{4}$
defined in section 2. 
In general we know that the monodromy group $\Gamma$
for these cases must satisfy $\Gamma(n)\subseteq\Gamma\subseteq\gn$
with $n=2,3,4$, respectively.
The key data are the orders of the zeros of
$g_2$, $g_3$ and $\delta=g_2^3-27 g_3^2$ in the  
associated Weierstrass form $y^2=4 x^3-g_2 x-g_3$
of the elliptic fibration. For the three 
curves we have  
\eqn\elldata{
\eqalign{\bX{2}{2} :
& \ \ \ g_2=f_4^2+12 f_8, \qquad g_3={1\over\sqrt{27}} (36 f_8-f_4^2) f_4, \cr 
& \ \ \ \delta=108 f_8 (4f_8-f_4^2)^2 \cr
\bX{2}{3}:  
& \ \ \ g_2=f_2(f_2^3+216 f_6),\quad 
g_3={1\over\sqrt{27}} (f_2^6-540f_6f_2^3-5832f_6^2),\cr 
& \ \ \ \delta=1728 f_6 (f_2^3-27 f_6)^3\cr
\bX{2}{4} :
& \ \ \ g_2=f_2^4+224 f_4 f_2^2+256f_4^2, \ \
g_3={1\over\sqrt{27}}(4096 f_4^3-8448 f_2^2 f_4^2-528 f_2^4f_4+f_2^6),\cr 
& \ \ \ \delta=1728f_2^2 f_4 (f_2^2-16f_4)^4\,.}}
{}From the above expression and Kodaira's results 
\ref\mon{K. Kodaira, {\sl On Compact Analytic Surfaces}, 
Annals of Mathematics, {\bf 77} (1963) 563.} 
as summarized in the first six columns 
of the table, we study the possible degenerations. The results are  
summarized in the last three columns. 
E.g. for $\IP^2_{1,1,2}[4]$ for a fiber of type $II$ to be present we need 
$\delta \sim t^2$ which is only possible if $i)$ 
$f_8\sim t^2$ or $ii)$ $4 f_8-f_4^2\sim t$. For case $i)$ the condition  
$g_3\sim t$ cannot be satisfied. Case $ii)$ requires $f_8\sim a+bt$
and $f_4\sim 4a+ct$ which contradicts the conditions on $g_2$ and $g_3$. 
Thus,the fiber type $II$ must be absent.

$${
\vbox{\offinterlineskip\tabskip=0pt
\halign{\strut
\vrule#&
~\hfil$#$~& 
\vrule$#$& 
~\hfil$#$~& 
\vrule$#$& 
~\hfil$#$~& 
\vrule$#$& 
~\hfil$#$~& 
\vrule$#$& 
~\hfil$#$~& 
\vrule$#$& 
~\hfil$#$~& 
\vrule$#$& 
~\hfil$#$~& 
\vrule$#$& 
~\hfil$#$~& 
\vrule$#$& 
~\hfil$#$~& 
\vrule#\cr
\noalign{\hrule}
&   {\rm o}(g_2) 
&&  {\rm o}(g_3)
&&  {\rm o}(\delta)
&&  {\rm fiber}
&&  {\rm singularity}
&&  {\rm mon} 
&&  \bX{2}{2}
&&  \bX{2}{3}
&&  \bX{2}{4}
&\cr
\noalign{\hrule}
&\ge 0&&\ge 0&&0&&smooth&& none&&\left(\matrix{1&0\cr 0&1}\right)&&   +&&+&&+& 
\cr
& 0   &&    0&&n&&I_n   && A_{n-1}&&\left(\matrix{1&n\cr 0&1}\right)&& 
{{1-8},\atop {10,\ldots ,24}} &&{1-6,\atop 9,\ldots, 24 }  &&{{1-4,6,8}\atop 
{12,\ldots. 
24}}& \cr
&\ge 1&&    1&&2&&II    && none   &&\left(\matrix{1&1\cr -1&0}\right)&& -&&-&&-& 
\cr
&\ge 1&&\ge 2&&3&&III   && A_1    &&\left(\matrix{0&-1\cr 1&0}\right)&& +&&-&&-& 
\cr
&\ge 2&&    2&&4&&IV    && A_2    &&\left(\matrix{0&1\cr -1&-1}\right)&& 
-&&+&&-& \cr
&    2&&\ge 3&&n+6&&I^*_n&&D_{n+4}&&-\left(\matrix{1&n\cr 0&1}\right)&& 
0-4&&-&&1& \cr
&\ge 2&&    3&&n+6&&I^*_n&&D_{n+4}&&-\left(\matrix{1&n\cr 0&1}\right)&&  
-&&-&&-& \cr
&\ge 3&&    4&&8&&IV^*   && E_6   &&\left(\matrix{-1&-1\cr 1&0}\right)&& 
-&&+&&-& \cr
&    3&&\ge 5&&9&&III^*  && E_7  &&\left(\matrix{0&-1\cr 1&0}\right)&&+&& -&&-& 
\cr
&\ge 4&&    5&&10&&II^*  && E_8   &&\left(\matrix{0&-1\cr 1&1}\right)&&-&& -&&-& 
\cr
\noalign{\hrule}}
\hrule}}$$ 
{}
This analysis allows us to derive the monodromy groups.
 
$\Gamma_0(2)$ has parabolic vertices and vertices 
of order 2 but none of order three (see e.g. \ref\lehner{J.\ J.\ Lehner, {\sl 
Discontinuous groups and automorphic functions}, Math. Surveys 8, 
AMS (1982), c1964.}).
The table shows that the $II,II^*,IV,IV^*$ fibers, which act on 
$\tau$ with order three, 
are indeed excluded. Of course the monodromies in table 1 result from a 
local analysis and are therefore determined only up to 
conjugation. However, from the fact that the 
generated monodromy $\Gamma$ has the property $\Gamma\subset SL(2,\IZ)$ 
and contains the shift $T$ as well as elements of 
order $2$ but no element of order $3$, we  
conclude that $\Gamma=\Gamma_0(2)$, up to conjugation.
If $f_8$ factorizes $f_8={\hat f}_4^2$, 
the Weierstrass form factorizes completely 
and odd $T$ shifts as well as order $2$ and $3$
elements are absent, which shows that the monodromy is $\Gamma(2)$.   

$\Gamma_0(3)$ has parabolic vertices and vertices 
of order three and as the generated
monodromy $\Gamma$ contains $T$ and elements of order 
$3$, but no element of order $2$, 
as can be seen from the table,  $\Gamma$ must be, 
again up to conjugation, $\Gamma_0(3)$. However, 
not all singular fibers which are 
compatible with $\Gamma_0(3)$ do actually 
occur. If $f_6$ factorizes as $f_6={\hat f}_2^3$, 
we see only $T^{3n}$ shifts  
and neither order 3 nor order 2 elements, hence we get back $\Gamma(3)$.

Finally $\Gamma_0(4)$ has only parabolic vertices and again from the table we
see that all finite order monodromies are indeed absent, 
while the $T$ shift is present, 
leading to the conclusion that the monodromy is $\Gamma_0(4)$. 
If $f_4=\hat f_2^2$, the Weierstrass form again completely factorizes. 
We get no finite order generators, no shift $T$, but
$T^2$ and hence a mondromy $\Gamma(4)$ $\subset \Gamma\subset$ $\Gamma_0(4)$.
$\Gamma(4)$ is only recovered if $f_4=f_1^4$ and $f_2={\hat f}_1^2$, in which 
all
complex moduli of the compactification are fixed.

\listrefs
\end